\documentclass[12pt,a4paper]{article}

\usepackage{a4wide}
\usepackage{amsmath}
\usepackage{bm}
\usepackage{amssymb}
\usepackage{hyperref}
\usepackage{epic}

\newcommand{\D}{d}
\newcommand{\BesselJ}{{J}}
\newcommand{\sign}{\mathrm{sign}}

\newcommand{\Tr}{{\rm Tr \,}}
\newcommand{\Op}{\mathcal{O}}
\newcommand{\fldZ}{\mathcal{Z}}
\newcommand{\fldX}{\mathcal{X}}
\newcommand{\fldY}{\mathcal{Y}}
\newcommand{\fldD}{\mathcal{D}}
\newcommand{\fldU}{\mathcal{U}}
\newcommand{\fldV}{\mathcal{V}}
\newcommand{\fldF}{\mathcal{F}}
\newcommand{\alg}[1]{\mathfrak{#1}}
\newcommand{\superN}{\mathcal{N}}

\newcommand{\osca}{\mathbf{a}}

\newcommand{\oscc}{\mathbf{c}}

\newcommand{\vac}{|0\rangle}
\newcommand{\density}{\hat \rho}
\newcommand{\auxdensity}{\hat \eta}

\makeatletter
\def\mr@ignsp#1 {\ifx\:#1\@empty\else #1\expandafter\mr@ignsp\fi}%
\newcommand{\multiref}[1]{\begingroup%\let\protect\string%
\xdef\mr@no@sparg{\expandafter\mr@ignsp#1 \: }%
\def\mr@comma{}%
\@for\mr@refs:=\mr@no@sparg\do{\mr@comma\def\mr@comma{,}\ref{\mr@refs}}%
\endgroup}
\makeatother

\newcommand{\Appref}[1]{Appendix~\multiref{#1}}

\numberwithin{equation}{section}

\begin{document}
\thispagestyle{empty}

\begin{flushright}\footnotesize
\texttt{hep-th/0702151}\\
\texttt{AEI-2007-008}\\
\vspace{0.5cm}
\end{flushright}
\setcounter{footnote}{0}

\begin{center}
{\Large{\bf Nesting and Dressing }}
\vspace{15mm}

{\sc Adam Rej, Matthias Staudacher and Stefan Zieme}\\[5mm]

{\it Max-Planck-Institut f\"ur Gravitationsphysik\\
     Albert-Einstein-Institut \\
     Am M\"uhlenberg 1, D-14476 Potsdam, Germany}\\[3mm]
     
\texttt{arej@aei.mpg.de}\\ 
\texttt{matthias@aei.mpg.de}\\ 
\texttt{stzieme@aei.mpg.de}\\[30mm]

\textbf{Abstract}\\[2mm]
\end{center}

\noindent{We compute the anomalous dimensions of field strength 
operators Tr$\,{\mathcal{F}^L}$ in $\mathcal{N}=4$ SYM from an 
asymptotic nested Bethe ansatz to all-loop order. Starting from the 
exact solution of the one-loop problem at arbitrary $L$,
we derive a single effective integral equation for the thermodynamic
$L \rightarrow \infty$ limit of these dimensions. 
We also include the recently proposed phase 
factor for the S-matrix of the planar AdS/CFT system. 
The terms in the effective equation corresponding to,
respectively, the nesting and the dressing are structurally
very similar. This hints at the physical origin of the
dressing phase, which we conjecture to arise from the hidden
presence of infinitely many auxiliary Bethe roots describing
a non-trivial ``filled'' structure of the theory's BPS 
vacuum. We finally show that the mechanism for creating
effective nesting/dressing kernels is quite generic by also
deriving the integral equation for the all-loop 
dimension of a certain one-loop $\mathfrak{so}(6)$
singlet state.  
}
%%%%%%%%%%%%%%%%%%%%%%%%%%%%%%%%%%%%%%%%%%%%%%%%%%%%%%%%%%%%%%%%%%%%%%
\newpage

\setcounter{page}{1}
%%%%%%%%%%%%%%%%%%%%%%%%%%%%%%%%%%%%%%%%%%%%%%%%%%%%%%%%%%%%%%%%%%%%%%
\section{Motivation, Conclusion and Overview}

There is much evidence that planar $\mathcal{N}=4$ SYM theory is
integrable and that its spectral problem is therefore exactly solvable. 
It was shown by Minahan and Zarembo
that the dilatation operator in the scalar matter sector at one 
loop can be mapped to the Hamiltonian of an integrable $\mathfrak{so}(6)$ 
spin chain and hence its eigenvalues can be found with a Bethe ansatz 
\cite{Minahan&Zarembo}. This extends to the full set of
operators, leading to an integrable non-compact 
nearest-neighbor super magnet \cite{super spin chain}.
The special ``solvable'' properties of the  $\mathcal{N}=4$ 
model under dilatation were already hinted at by Lipatov in 
\cite{Lip}, and extend a rather generic if incomplete phenomenon 
in more general gauge theories such as QCD, 
as first shown by Belitsky, Braun, Derkachov, Korchemsky 
and Manashov \cite{Bel}. 

The concept of factorized scattering, one of the hallmarks of 
integrability, can be extended to higher loop orders \cite{Factorized S-matrix},
and to strong coupling \cite{AFS}, where the gauge theory
is expected to be more suitably described by a superstring
theory in a curved $AdS_5 \times S^5$ background.
Various analyses of this topic 
\cite{various analyses} led to a set of {\it asymptotic} all-loop 
Bethe equations for the full theory \cite{LongRange}. 
It should be stressed that quantum integrability remains to
be {\it proven} in both gauge and string theory. In particular,
on the gauge side one phenomenologically finds an
integrable long-range spin chain, and the all-loop
factorization of the multi-body magnon S-matrix into two-body 
processes currently has to be {\it assumed}. Correspondingly,
symmetry fixes the magnon S-matrix only up to an overall phase 
factor \cite{Factorized S-matrix,LongRange,su(2|2) s-matrix}.
The latter encodes our lack of understanding of the 
underlying microscopic integrable structure\footnote{
As shown in \cite{su(2|2) s-matrix}, the full
S-matrix, which is proportional to the tensor product of two
copies of Shastry's R-matrix for the Hubbard model, 
satisfies the Yang-Baxter equation (see also
\cite{Arutyunov:2006yd}).
It should be stressed, however, that Yang-Baxter symmetry of the 
{\it two-}body S-matrix
is a necessary, but not sufficient
condition for the integrability of a given system. An infamous 
counter-example is the bosonic Hubbard model.
}.
However, as was argued by Janik, the dressing phase may
be constrained by invoking crossing-invariance\footnote{
Crossing symmetry should also be a consequence of --
rather than an axiom for -- the 
proper microscopic formulation of the planar AdS/CFT system.
} 
\cite{Janik}.
And indeed the string S-matrix satisfies crossing to the known
\cite{AFS,Hernandez:2006tk} orders \cite{Arutyunov:2006iu}.
A proposal for the complete structure of the 
dressing phase has recently been made in  \cite{TransCross} by
combining Bethe ansatz techniques for the all-order perturbative
large spin limit of  Wilson twist operators \cite{InteTrans}
with conjectures by 
Beisert, Hern\'andez and L\'opez \cite{Beisert:2006ib}
on the full (asymptotic) structure of the string Bethe ansatz \cite{AFS}.

An independent four-loop calculation strongly supports this guess \cite{Bern}. This calculation scheme has been algorithmically 
improved and leads to agreement between field theory 
and the dressed Bethe ansatz with a margin of error of
$0.001 \%$ \cite{Cachazo:2006az}.
Incorporating this dressing phase into the Bethe ansatz, an integral 
equation for the universal scaling function $f(g)$ of $\mathcal{N}=4$ 
gauge theory in the large spin limit was derived in \cite{TransCross}. 
Some of the analytic properties of the solution of this equation were 
analyzed in \cite{highest trans}, and in the leading strong coupling limit 
the behavior of the scaling function as predicted by string theory 
\cite{Gubser:2002tv} through the AdS/CFT correspondence 
 agrees with the analysis of the equation 
 \cite{Benna:2006nd,Alday:2007qf,Kostov:2007kx,Beccaria:2007tk}. 
The subleading order was also successfully, albeit numerically,
compared in \cite{Benna:2006nd} to the known string result
\cite{Frolov:2002av}.

The dressing phase of \cite{TransCross} takes a surprisingly
complex and seemingly opaque form. Arguing that it should be
of a fundamental nature is, to put it mildly, unconvincing. Let
us recall its structure. The phase shift 
$2\, \theta(u,u')=-i\,\log\, \sigma^2(u,u')$ stemming from the
dressing factor $\sigma^2(u,u')$ when two magnons
with rapidities $u$,$u'$ pass each other (see \cite{LongRange,TransCross} 
for the notation) is 
\begin{equation}\label{dressingfactor}
2\,i\,\theta(u,u')=2\,g^2\,
\int_{-\infty}^{\infty} dt\,e^{i\,t\,u}\,e^{-\frac{|t|}{2}}
\int_{-\infty}^{\infty} dt'\,e^{i\,t'\,u'}\,e^{-\frac{|t'|}{2}}\,
\left(\hat{K}_{d}(2\,g\,t,2\,g\,t')-\hat{K}_{d}(2\,g\,t',2\,g\,t)\right)\, ,
\end{equation}
where the ``magic kernel'' reads
\begin{equation}\label{Kd}
\hat{K}_{d}(t,t')=
8\,g^{2}\int^{\infty}_{0}\D t'' \hat{K}_{1}(t,2\,g\,t'')\,
\frac{t''}{e^{t''}-1}\,\hat{K}_{0}(2\,g\,t'',t')\, ,
\end{equation}
and the symmetric kernels $\hat K_0$, $\hat K_1$ are
expressed with the help of Bessel functions as
\begin{eqnarray}
\hat{K}_{0}(t,t')&=&\frac{t\,\BesselJ_{1}(t)\,\BesselJ_{0}(t')-
t'\,\BesselJ_{0}(t)\,\BesselJ_{1}(t')}{t^{2}-t'^{2}}\,,\label{K0}\\
\hat{K}_{1}(t,t')&=&\frac{t'\,\BesselJ_{1}(t)\,\BesselJ_{0}(t')-
t\,\BesselJ_{0}(t)\,\BesselJ_{1}(t')}{t^{2}-t'^{2}}\label{K1}\,.
\end{eqnarray}
This way of writing the dressing phase has a decidedly
``thermodynamic'' flavor. Nevertheless, the claim of \cite{TransCross}
is certainly that this phase should contribute to the anomalous
dimensions of {\it short} operators\footnote{
Strictly speaking the phase was found in \cite{TransCross}
by studying the large-spin limit of twist operators. This
is very similar to a thermodynamic limit in that it also
involves infinitely many Bethe roots \cite{InteTrans}.
It would be very important to check the phase at four loops 
for a finite, short operator without taking any limit.
}
at and beyond four-loop order, up until ``wrapping order''.
A natural explanation of its structure could come from a
hidden ``non-trivial'' vacuum. Recall that this is usually the case in
relativistic field theories, or else, in ``physical''
ground state configurations of magnetic systems such
as the Heisenberg or Hubbard antiferromagnets.
What we are saying is that the
dressing factor indicates that the BPS states 
\begin{equation}\label{bps}
\Tr \fldZ^L\, ,
\end{equation}
do {\it not} constitute boring
``ferromagnetic'' reference vacua, but are rather states more akin to an  
``antiferromagnetic'', ``physical'' vacuum. For reasons to
be understood, BPS vacuum polarization at weak coupling
appears for the first time at four-loop order. The difference,
in the $\mathfrak{su}(2)$ sector, of the AdS/CFT system to
the Hubbard model \cite{hubbard} is then that in the latter this 
vacuum polarization does not appear. Toy models approximatively
demonstrating a similar phenomenon in the present context have 
appeared previously in \cite{mann-polchinski, gromov-kazakov}. 

In the current paper we will not rigorously 
analyze the
underlying vacuum structure of the BPS states \eqref{bps}, 
which requires to go beyond the Bethe equations of \cite{LongRange}. 
In order to nevertheless
prove our point, we will instead focus on
another ``false vacuum", but one which may be treated with the
asymptotic spectral equations in \cite{LongRange}.
Namely, we will consider operators which at one loop are built from a 
sole field strength component  $\mathcal{F}$:
\begin{equation}\label{ops}
\Tr \mathcal{F}^L\, .
\end{equation}
These are of interest as they are not embedded in a rank-one subsector, 
and are therefore described by an asymptotic 
{\it nested} Bethe ansatz. If $L$ is large enough, 
i.e.~the operator is ``long'',
we may trust the ansatz to arbitrary loop order. In fact, we
may pass to the thermodynamic limit $L \rightarrow \infty$, 
in which the dimension of the operators \eqref{ops} becomes
\begin{equation}\label{dim}
\Delta(g)=2\,L+
8\,L\,g^2\,\int^{\infty}_{0}\D t~\frac{\BesselJ_{1}(2\,g\,t)}{2\,g\,t}\,
\density(t)\, ,
\end{equation}
where $\density(t)$ is essentially the Fourier-transform of
the density of momentum-carrying first-level Bethe roots.
As is common in thermodynamic situations, the non-linear
nested Bethe equations may be turned into a system
of linear integral equations for $\density(t)$ in conjunction
with a number of further auxiliary densities describing the 
distribution of the higher-level Bethe roots. Interestingly, the
auxiliary densities may be eliminated, and one ends up with
a single, {\it effective} linear equation for
the principal density $\density(t)$. It reads
\begin{eqnarray}\label{finaleq}
\density(t)&=&e^{-2t}\,(1+
e^{t})\Big [ \BesselJ_{0}(2\,g\,t)\nonumber\\
&&-4\,g^{2}\,t\int^{\infty}_{0}\D t'\left ( \frac{e^{t}}{e^{t}+1}\,
\hat{K}_{0}(2\,g\,t,2\,g\,t')+
\frac{e^{t'}}{e^{t'}+1}\,\hat{K}_{1}(2\,g\,t,2\,g\,t')\right)
\density(t')\nonumber\\
&&-4\,g^{2}\,t\,\int^{\infty}_{0}\D t'\left (
\hat{K}_n(2\,g\,t,2\,g\,t')+\hat{K}_{d}(2\,g\,t,2\,g\,t')\right)\density(t')\Big]\,.\nonumber\\
\end{eqnarray}
Here $\hat K_d$ is the above kernel \eqref{Kd} summarizing 
the effects of the {\it dressing} phase. Interestingly, together with a 
few further terms in the second line of \eqref{finaleq},
the net influence of the {\it nesting} is exerted by the kernel
$\hat{K}_n$ which reads
\begin{equation}\label{Kn}
\hat{K}_n(t,t')=4\,g^{2}\,\int^{\infty}_{0}\D t'' \hat{K}_{1}(t,2\,g\,t'')\,\frac{t''}{e^{t''}+1}\,\hat{K}_{0}(2\,g\,t'',t')\,.
\end{equation}
We are confident that the reader's sharp eyes will spot
the structural similarity between this nesting kernel 
$\hat{K}_n$ and the dressing kernel $\hat K_d$ in \eqref{Kd}.
In conclusion, this result suggests that the AdS/CFT phase factor 
\cite{TransCross} is generated by hidden, nested levels of a 
yet-to-be constructed final Bethe ansatz.

In the remainder of this paper we will work out the asymptotic 
anomalous dimensions of the operators \eqref{ops} and derive
the above equations. The detailed calculations are rather technical,
and proceed as follows: In section \ref{sec:setup} we 
obtain the asymptotic spectrum of these operators,
which form a long-range spin chain, by applying the
Bethe equations of \cite{LongRange}.  In section \ref{sec:one-loop}
we solve the one-loop problem, i.e.~we find the principal and
auxiliary one-loop Bethe roots for the appropriate Dynkin diagram.
This solution is exact for all finite values of the length $L$.
We also find all one-loop higher conserved charges at arbitrary $L$.
The heart of the paper is section \ref{sec:thermodynamic},
where we work out the thermodynamic limit of our one-loop solution, 
find (thermodynamically) all higher-loop perturbations,
and reduce the resulting system of integral equations to the single
effective one in \eqref{finaleq}. 
Finally, in section \ref{sec:SO6singlet}, we
study a few further thermodynamic 
high-energy states satisfying special filling conditions. 
Apart from exhibiting certain intriguing transcendentality properties,
the examples provide further support for our picture of the origin of
nesting and dressing factors. We end with a short outlook.
As always when working with Bethe equations,
great care has to be taken to understand the distribution of
roots. The appendices contain numerous checks and considerations
justifying the methodology and statements in the main body of the text.

%%%%%%%%%%%%%%%%%%%%%%%%%%%%%%%%%%%%%%%%%%%%%%%%%%%%%%%%%%%%%%%%%%%%%%
\section{Excitation Scheme for Field Strength Operators}
\label{sec:setup}

We are interested in the pseudo-vacuum states
$\Tr \mathcal{F}^L$ in \eqref{ops}. 
Here pseudo-vacuum refers to the fact that
these states may serve as a reference vacuum for the 
one-loop Bethe ansatz \cite{super spin chain}.
In an oscillator realization of the system, see  \cite{Niklas thesis},
$\mathcal{F}^{L}$ is the tensor product of $L$ fields 
$|\mathcal{F}\rangle$ where
\begin{equation}\label{F}
|\mathcal{F}\rangle=\mathbf{a}_{1}^{\dagger}\mathbf{a}_{1}^{\dagger}|0\rangle\, ,
\end{equation}
and $\mathbf{a}_{1}^{\dagger}$ are bosonic creation operators. 
After imposing the trace condition, these states are gauge-invariant,
but, in contradistinction to the superficially similar
BPS states $\Tr \fldZ^L$ in \eqref{bps}, not protected.
In fact, their anomalous dimension is very large \cite{Niklas thesis}. 
This type of operator was studied at one-loop in an 
integrable sector of QCD in \cite{ferretti}.

There exist alternative choices for the Dynkin diagrams of a 
superalgebra due to the freedom to choose each node to be
either fermionic or bosonic, with the constraint that at least 
one node has to be fermionic. 
Changing from one diagram to another changes the
excitation pattern of the spin chain, and sometimes 
even induces a change of the corresponding vacuum state. 
In particular, the one-loop 
vacuum of the standard distinguished Dynkin diagram 
of $\mathfrak{psu}(2,2|4)$ is precisely given by the 
states $\Tr \mathcal{F}^L$ \cite{super spin chain}.
Therefore no Bethe ansatz, which takes care of
the diagonalization of excitations, is necessary, and
the one-loop anomalous dimension is just the vacuum-energy.
However, it is not known how to deform the one-loop Bethe
equations of the distinguished Dynkin diagram to higher
loops. The same is true for most other choices, and
the long-range Bethe equations of \cite{LongRange}
only appear to ``work'' for a very specific diagram.
In that diagram the vacuum is $\Tr \fldZ^L$, and
$\Tr \mathcal{F}^L$ is a highly excited state with
many excitations, whose momenta have to be diagonalized
in order to reproduce the correct energy.

A somewhat similar pseudo-vacuum consists of a cyclic
tensor product of $L$ fermions  $\fldU$, 
i.e.~$\Tr \fldU^L$, which was studied in \cite{HES SU(1|1)}.
However, while the latter lies in a rank-one subsector of the
symmetry group,  $\mathfrak{su}(1|1)$, the states
$\Tr \mathcal{F}^L$ are not confined to any such subsector.
Correspondingly, in order to diagonalize the latter
one needs a {\it nested} Bethe ansatz.
The excitation pattern of Bethe roots for the higher-loop
Dynkin diagram of \cite{LongRange} reads
\begin{equation}\label{pattern}
(K_1,K_2,K_3,K_4,K_5,K_6,K_7)=(0,0,2L-3,2L-2,L-1,L-2,L-3)\, ,
\end{equation}
where the $K_\nu$ is the excitation number of the $\nu$-th node
of the Dynkin diagram.
Clearly all but the first two nodes are highly excited.

\begin{center}
\begin{figure}[t]\label{magic square}
\begin{minipage}{360pt}
\setlength{\unitlength}{1pt}
\small\thicklines
\begin{center}
\begin{picture}(200,200)(30,-180)
\put( 55,-144){\makebox(0,0)[t]{$\mathbf{c}^\dagger_1 \mathbf{a}^{2}$}}
\put( 80,-152){\circle{15}}
\put( 80,-119){\line(0,-1){26}}
\put( 55,-104){\makebox(0,0)[t]{$\mathbf{a}^\dagger_2 \mathbf{a}^{1}$}}
\put( 80,-112){\circle{15}}
\put( 80,-79){\line(0,-1){26}}
\put( 55,-64){\makebox(0,0)[t]{$\mathbf{a}^\dagger_1 \mathbf{c}^{2}$}}
\put( 80,-72){\circle{15}}
\put( 80,-07){\line(0,-1){58}}
%%%
\put( 80,00){\circle{15}}
\put( 80,25){\makebox(0,0)[t]{$\mathbf{c}^\dagger_2 \mathbf{d}^\dagger_2$}}
\put( 87,00){\line(1,0){66}}
\put(120,-35){\makebox(0,0)[b]{$\fldX$}}
\put(120,-75){\makebox(0,0)[b]{$\dot{\fldU}$}}
\put(120,-115){\makebox(0,0)[b]{$\dot{\fldV}$}}
\put(120,-155){\makebox(0,0)[b]{$\bar{\fldY}$}}
\put(160,00){\circle{15}}
\put(160,25){\makebox(0,0)[t]{$\mathbf{d}^2 \mathbf{b}^\dagger_1$}}
\put(160,-35){\makebox(0,0)[b]{$\fldU$}}
\put(160,-75){\makebox(0,0)[b]{$\fldD$}}
\put(160,-115){\makebox(0,0)[b]{$\dot{\fldD}$}}
\put(160,-155){\makebox(0,0)[b]{$\bar{\fldV}$}}
\put(167,00){\line(1,0){26}}
\put(200,00){\circle{15}}
\put(200,25){\makebox(0,0)[t]{$\mathbf{b}^1 \mathbf{b}^\dagger_2$}}
\put(200,-35){\makebox(0,0)[b]{$\fldV$}}
\put(200,-75){\makebox(0,0)[b]{$\dot{\bar{\fldD}}$}}
\put(200,-115){\makebox(0,0)[b]{$\bar{\fldD}$}}
\put(200,-155){\makebox(0,0)[b]{$\bar{\fldU}$}}
\put(207,00){\line(1,0){26}}
\put(240,00){\circle{15}}
\put(240,25){\makebox(0,0)[t]{$\mathbf{b}^2 \mathbf{d}^\dagger_1$}}
\put(240,-35){\makebox(0,0)[b]{$\fldY$}}
\put(240,-75){\makebox(0,0)[b]{$\dot{\bar{\fldV}}$}}
\put(240,-115){\makebox(0,0)[b]{$\dot{\bar{\fldU}}$}}
\put(240,-155){\makebox(0,0)[b]{$\bar{\fldX}$}}
\put( 75,-157){\line(1,1){10}}
\put( 75,-147){\line(1,-1){10}}
\put( 75,-77){\line(1,1){10}}
\put( 75,-67){\line(1,-1){10}}
\put(155,-5){\line(1,1){10}}
\put(155,5){\line(1,-1){10}}
\put(235,-5){\line(1,1){10}}
\put(235,5){\line(1,-1){10}}
%%%
\linethickness{0.75pt}
\multiput(100,-15)(40,0){5}{\line(0,-1){160}}
\multiput(100,-15)(0,-40){5}{\line(1,0){160}}
\end{picture}
\end{center}
\end{minipage}
\caption{Oscillator realization and fundamental magnons.
There are four adjoint complex scalars
$\fldX,\bar \fldX,\fldY,\bar \fldY$,
four light-cone covariant derivatives, 
$\fldD, \bar \fldD, \dot \fldD, \dot{\bar \fldD}$,
and eight adjoint fermions 
$\fldU,\fldV,\dot \fldU,\dot \fldV,\bar \fldU,\bar \fldV,
\dot{\bar \fldU},\dot{\bar \fldV}.$
The nodes of the Dynkin diagram are labeled from
1 to 7, starting at the south-west end and 
going clockwise until  the north-east end is reached.
}
\end{figure}
\end{center}

It is interesting to understand the state $\Tr \mathcal{F}^L$ 
in the picture of the elementary excitations of the 
long-range spin chain: the 8+8 magnons of the
$\mathcal{N}=4$ model.
We tabulated in Fig.~1 all ``fundamental magnons'' 
in a way which allows to read of the excitation pattern in
the higher-loop Dynkin diagram with ease.
We also included the magnon-creation operators
in the oscillator picture, cf.~\cite{Niklas thesis}.
The $\alg{su}(2|2) \oplus \alg{su}(2|2)$ invariant
S-matrix acts, respectively, on the rows and columns of
Fig.~1.
Exciting the central 4-th node $K_4$ times
inserts $K_4$ $\fldX$ bosons into the BPS vacuum $\fldZ^L$.
The $K_4$ Bethe roots of the central node parametrize
the momenta of these magnons. Next, exciting $K_5$
times the fifth node converts $K_5$ of the $\fldX$ bosons
into $\fldU$ fermions. The process continues until
all magnons contained in a given state are created.
It follows from \eqref{pattern}
that the fields $\mathcal{F}$ in $\Tr \mathcal{F}^L$ 
are not elementary. Instead, each $\mathcal{F}$ is 
essentially equivalent to a composite of two
fermions $\dot \fldU$ and $\dot{\bar{\fldV}}$ 
{\it minus} one field $\fldZ$. Let us explain this in more
detail for a single $\fldF$. The two fermions and the
background field $\fldZ$ are
expressed in terms of the oscillators as
\begin{equation}
 \dot{\fldU}=\osca_1^\dagger\oscc_4^\dagger \vac\, \quad \dot{\bar{\fldV}}=\osca_1^\dagger\oscc_3^\dagger\vac\,
 \quad \fldZ=\oscc_3^\dagger\oscc_4^\dagger\vac\,.
\end{equation}
and form a $\fldF$, {\it cf}~\eqref{F}, while inserting another $\fldZ$:
\begin{equation}
\dot{\fldU}\,\dot{\bar{\fldV}}=
\,\osca_{1,1}^\dagger \oscc_{4,1}^\dagger \, \osca_{1,2}^\dagger \oscc_{3,2}^\dagger \,
 \vac_1 \otimes \vac_2 
\quad \leftrightarrow \quad 
\fldF\, \fldZ = \osca_{1,1}^\dagger \osca_{1,1}^\dagger \, \oscc_{3,2}^\dagger\oscc_{4,2}^\dagger\,
\vac_1 \otimes \vac_2\,.
\end{equation}
We see that each $\fldF$ corresponds to two magnon
excitations $\dot{\fldU}\,\dot{\bar{\fldV}}$.
Removing $L$ background fields $\fldZ$ does not
change the excitation number, but reduces the length
(as well as the R-charge) by $L$ units.
Correspondingly, the state $\Tr \mathcal{F}^L$
has, at one loop, $2 L$ excitations living on a lattice of $L$ sites.
Beyond one loop, this length is not conserved.

In order to write down the Bethe equations it is useful
to define spectral parameters $x^\pm,u^\pm,u$ which
are interrelated by
\begin{equation}\label{definition x}
x(u)=\frac{u}{2}\left(1+\sqrt{1-4\,\frac{g^2}{u^2}}\right), \qquad 
x^{\pm}=x(u^\pm)\, , \qquad
u^\pm=u\pm\tfrac{i}{2}\, ,
\end{equation}
and the coupling constant $g$ is proportional to
the square root of the `t~Hooft coupling constant
\begin{equation}
g=\frac{\sqrt{\lambda}}{4\pi}\, .
\end{equation}
These parameters are assigned double indices, 
i.e.~the Bethe equations are written in terms of
$x^\pm_{\nu,k}$, $u^{ }_{\nu,k}$ where $\nu=1,\ldots,7$ denotes
the nesting level (or node-number of the Dynkin diagram),
and $k=1,\ldots,K_\nu$ labels the Bethe roots of level $\nu$.

The Bethe equations for the specific excitation pattern
\eqref{pattern} are given by ({\it cf} \cite{LongRange})
\begin{eqnarray}\label{starting BE}
1&=&\prod_{j=1}^{2L-2}\frac{x_{3,k}-x_{4,j}^{+}}{x_{3,k}-x_{4,j}^{-}}\nonumber\\
\left (\frac{x_{4,k}^{+}}{x_{4,k}^{-}} \right )^{L}&=&\prod_{\substack{j=1 \\j\neq k}}^{2L-2}\frac{x_{4,k}^{+}-x_{4,j}^{-}}{x_{4,k}^{-}-x_{4,j}^{+}}\frac{1-g^{2}/x_{4,k}^{+}x_{4,j}^{-}}{1-g^{2}/x_{4,k}^{-}x_{4,j}^{+}}\,
\sigma^2(u_{4,k},u_{4,j})
\nonumber\\
&&\times\prod_{j=1}^{2L-3}\frac{x_{4,k}^{-}-x_{3,j}}{x_{4,k}^{+}-x_{3,j}} \prod_{j=1}^{L-1}\frac{x_{4,k}^{-}-x_{5,j}}{x_{4,k}^{+}-x_{5,j}}\prod_{j=1}^{L-3}\frac{1-g^{2}/x_{4,k}^{-}x_{7,j}}{1-g^{2}/x_{4,k}^{+}x_{7,j}}\,
\nonumber\\
1&=&\prod_{j=1}^{L-2}\frac{u_{5,k}-u_{6,j}+\frac{i}{2}}{u_{5,k}-u_{6,j}-\frac{i}{2}}\prod_{j=1}^{2L-2}\frac{x_{5,k}-x_{4,j}^{+}}{x_{5,k}-x_{4,j}^{-}}\nonumber\\
1&=&\prod_{\substack{j=1\\j \neq k}}^{L-2}\frac{u_{6,k}-u_{6,j}-i}{u_{6,k}-u_{6,j}+i}\prod_{j=1}^{L-1}\frac{u_{6,k}-u_{5,j}+\frac{i}{2}}{u_{6,k}-u_{5,j}-\frac{i}{2}}\prod_{j=1}^{L-3}\frac{u_{6,k}-u_{7,j}+\frac{i}{2}}{u_{6,k}-u_{7,j}-\frac{i}{2}}\nonumber\\
1&=&\prod_{j=1}^{L-2}\frac{u_{7,k}-u_{6,j}+\frac{i}{2}}{u_{7,k}-u_{6,j}-\frac{i}{2}}\prod_{j=1}^{2L-2}\frac{1-g^{2}/x_{7,k}x_{4,j}^{+}}{1-g^{2}/x_{7,k}x_{4,j}^{-}},
\end{eqnarray}
where the dressing factor $\sigma^2(u_k,u_j)=e^{2 i \theta(u_k,u_j)}$ 
is given by \eqref{dressingfactor}.

The (asymptotic) anomalous dimensions of an operator can 
then be deduced from the energy $E(g)$  
of the long-range spin chain by
\begin{equation}\label{dimeng}
\Delta=\Delta_0+2\,g^2\, E(g)\,,
\end{equation}
where $\Delta_0$ denotes the operator's classical dimension. 
In the case of the field strength operators \eqref{ops} it is given by 
$\Delta_0=2L$. In turn, this energy is found from the
``momentum-carrying'' Bethe roots living on the central (i.e.~4-th)
node as
\begin{equation}\label{chainenergy}
E(g)=\sum_{k=1}^{2L-2}\left (\frac{i}{x_{4,k}^{+}}-\frac{i}{x_{4,k}^{-}} \right )\, .
\end{equation}

Let us note that the state  $\dot{\mathcal{F}}^{L}$, which is a tensor product of
\begin{equation}
|\dot{\mathcal{F}}\rangle=\mathbf{b}_{1}^{\dagger}\mathbf{b}_{1}^{\dagger}\mathbf{c}_{1}^{\dagger}\mathbf{c}_{2}^{\dagger}\mathbf{c}_{3}^{\dagger}\mathbf{c}_{4}^{\dagger}|0\rangle\,,
\end{equation}
has mirror-inverted excitation numbers and hence the same Bethe equations \eqref{starting BE}.

We will firstly solve the one loop-problem at finite $L$, 
and secondly construct the thermodynamic limit $L \to \infty$. 
Subsequently we will focus on the all-loop equations and 
derive an integral equation for the corresponding root density.

%%%%%%%%%%%%%%%%%%%%%%%%%%%%%%%%%%%%%%%%%%%%%%%%%%%%%%%%%%%
\section{The One-Loop Problem}
\label{sec:one-loop}

The one-loop Bethe equations for the field strength operators 
$\Tr \mathcal{F}^L$ are obtained from the asymptotic
all-loop equations \eqref{starting BE} by taking the limit 
$g \to 0$. This yields
\begin{eqnarray}\label{oneloopbethe}
1&=&\prod^{2L-2}_{j=1} \frac{u_{3,k}-u_{4,j}-\frac{i}{2}}{u_{3,k}-u_{4,j}+\frac{i}{2}}\nonumber\\
\left(\frac{u_{4,k}+\frac{i}{2}}{u_{4,k}-\frac{i}{2}}\right)^L&=&\prod^{2L-2}_{\substack{j=1 \\j\neq k}} \frac{u_{4,k}-u_{4,j}+i}{u_{4,k}-u_{4,j}-i} \prod^{2L-3}_{j=1} \frac{u_{4,k}-u_{3,j}-\frac{i}{2}}{u_{4,k}-u_{3,j}+\frac{i}{2}} \prod^{L-1}_{j=1} \frac{u_{4,k}-u_{5,j}-\frac{i}{2}}{u_{4,k}-u_{5,j}+\frac{i}{2}}\nonumber\\
1&=&\prod^{L-2}_{j=1} \frac{u_{5,k}-u_{6,j}+\frac{i}{2}}{u_{5,k}-u_{6,j}-\frac{i}{2}}\prod^{2L-2}_{j=1} \frac{u_{5,k}-u_{4,j}-\frac{i}{2}}{u_{5,k}-u_{4,j}+\frac{i}{2}}\nonumber\\
1&=&\prod^{L-2}_{\substack{j=1 \\ j\neq k}} \frac{u_{6,k}-u_{6,j}-i}{u_{6,k}-u_{6,j}+i} \prod^{L-1}_{j=1} \frac{u_{6,k}-u_{5,j}+\frac{i}{2}}{u_{6,k}-u_{5,j}-\frac{i}{2}} \prod^{L-3}_{j=1} \frac{u_{6,k}-u_{7,j}+\frac{i}{2}}{u_{6,k}-u_{7,j}-\frac{i}{2}}\nonumber\\
1&=&\prod^{L-2}_{j=1} \frac{u_{7,k}-u_{6,j}+\frac{i}{2}}{u_{7,k}-u_{6,j}-\frac{i}{2}}\, .
\end{eqnarray}
These can be further simplified with a suitable dualization, 
see e.g.~\cite{CompleteSpectrum} and references therein.
We start by introducing the polynomial $P(u)$ for the $u_3$ roots as
\begin{equation}
P(u)=\prod^{2L-2}_{j=1} (u-u_{4,j}+\tfrac{i}{2})-\prod^{2L-2}_{j=1}(u-u_{4,j}-\tfrac{i}{2})=i (2L-2) \prod^{2L-3}_{j=1}(u-u_{3,j})\, .
\nonumber
\end{equation}
It is straightforward to find
\begin{equation}
\frac{P(u_{4,k}+\frac{i}{2})}{P(u_{4,k}-\frac{i}{2})}=\prod^{2L-2}_{\substack{j=1\\ j\neq k}}\frac{u_{4,k}-u_{4,j}+i}{u_{4,k}-u_{4,j}-i}=\prod^{2L-3}_{j=1}\frac{u_{4,k}-u_{3,j}+\frac{i}{2}}{u_{4,k}-u_{3,j}-\frac{i}{2}}\, .\nonumber
\end{equation}
Applying the same procedure to the roots $u_{7}$ and $u_{6}$ we reach 
an effective system of equations, namely
\begin{equation} \label{eff1}
\left(\frac{u_{4,k}+\frac{i}{2}}{u_{4,k}-\frac{i}{2}}\right)^L=\prod^{L-1}_{j=1} \frac{u_{4,k}-u_{5,j}-\frac{i}{2}}{u_{4,k}-u_{5,j}+\frac{i}{2}}
\end{equation}
\begin{equation} \label{eff2}
1=\prod^{L-1}_{\substack{j=1 \\ j\neq k}} \frac{u_{5,k}-u_{5,j}+i}{u_{5,k}-u_{5,j}-i} \prod^{2L-2}_{j=1} \frac{u_{5,k}-u_{4,j}-\frac{i}{2}}{u_{5,k}-u_{4,j}+\frac{i}{2}}\,.
\end{equation}
By introducing the polynomial $Q_{4}(u)$ for the momentum carrying roots $u_4$ as
\begin{eqnarray}\label{Q_4(u)}
Q_{4}(u)&=&(u+\tfrac{i}{2})^L \prod^{L-1}_{j=1}(u-u_{5,j}+\tfrac{i}{2})-(u-\tfrac{i}{2})^L \prod^{L-1}_{j=1}(u-u_{5,j}-\tfrac{i}{2})\nonumber\\
&=&i(2L-1)\prod^{2L-2}_{j=1}(u-u_{4,j})
\end{eqnarray}
and noting that
\begin{displaymath}
\frac{Q_{4}(u_{5,k}+\frac{i}{2})}{Q_{4}(u_{5,k}-\frac{i}{2})}=\left(\frac{u_{5,k}+i}{u_{5,k}-i}\right)^L\prod^{L-1}_{\substack{j=1 \\ j\neq k}}\frac{u_{5,k}-u_{5,j}+i}{u_{5,k}-u_{5,j}-i}=\prod^{2L-2}_{j=1}\frac{u_{5,k}-u_{4,j}+\frac{i}{2}}{u_{5,k}-u_{4,j}-\frac{i}{2}}\,,
\end{displaymath}
we find that (\ref{eff2}) simplifies enormously to the
free equation
\begin{equation}\label{decoupled u5}
\left(\frac{u_{5,k}+i}{u_{5,k}-i}\right)^L=1\,.
\end{equation}
This is solved by
\begin{displaymath}
u_{5,k}=\textrm{cot}\left(\frac{\pi k}{L}\right)\,, \qquad k=1,\ldots,L-1\,.
\end{displaymath}
Plugging this solution back into equation 
(\ref{eff1}) we infer that $u_{4,k}$ are roots of the equation
\begin{equation}
\bigg(\frac{u+\frac{i}{2}}{u-\frac{i}{2}}\bigg)^L=\prod^{L-1}_{j=1} \frac{u-\textrm{cot}(\frac{\pi j}{L})-\frac{i}{2}}{u-\textrm{cot}(\frac{\pi j}{L})+\frac{i}{2}}\,.
\end{equation}
On the other hand, the 
$u_5$ roots are zeros of the polynomial $Q_{5}(u)$ 
\begin{equation}\label{Q_5}
Q_{5}(u)=(u+i)^{L}-(u-i)^{L}\, ,
\end{equation}
such that we can rewrite (\ref{Q_4(u)}) as
\begin{equation}
Q_{4}(u)=(u+\frac{i}{2})^{L}\,Q_{5}(u+\frac{i}{2})-(u-\frac{i}{2})^{L}\,Q_{5}(u-\frac{i}{2})\,.
\end{equation}
We have therewith derived a polynomial $Q_{4}(u)$ of degree $2L-2$ 
\begin{equation}\label{Q}
Q_{4}(u)= (u+\tfrac{i}{2})^L 
\left((u+\tfrac{3}{2}i)^L-(u-\tfrac{1}{2}i)^L\right)+
(u-\tfrac{i}{2})^L \left((u-\tfrac{3}{2}i)^L-(u+\tfrac{1}{2}i)^L\right),
\end{equation}
whose zeros correspond to the {\it exact}
one-loop roots $u_{4}$ for {\it arbitrary, finite} $L$. All roots of 
this polynomial are real (for a proof, see \Appref{app:polyprop}).
%%%%%%%%%%%%%%%%%%%%%%%%%%%%%%%%%%%%%%%%%%%%%%%%%%%%%%%%%%%%%%
\subsection{Energy and Higher Conserved Charges}
Since the $u_{4,k}$ are the roots of (\ref{Q}) we can write
\begin{displaymath}
Q_{4}(u)=-2L(2L-1) \prod^{2L-2}_{j=1}(u-u_{4,j})\,.
\end{displaymath}
This allows us to express the energy 
$E_{\mathcal{F}^L}(g)=E_{\mathcal{F}^L}+\Op(g^2)$
\eqref{chainenergy}
in the one-loop approximation $E_{\mathcal{F}^L}$ 
in terms of $Q_{4}(u)$ 
\begin{equation}
E_{\mathcal{F}^L}=\sum^{2L-2}_{j=1} \frac{1}{\frac{1}{4}+u_{4,j}^2}=i\left(\frac{Q_{4}'(\frac{i}{2})}{Q_{4}(\frac{i}{2})}-\frac{Q_{4}'(-\frac{i}{2})}{Q_{4}(-\frac{i}{2})}\right)\,.
\end{equation}
On the other hand, using the explicit form (\ref{Q}) we derive
\begin{displaymath}
\frac{Q_{4}'(\frac{i}{2})}{Q_{4}(\frac{i}{2})}=-\frac{3}{2}\,i\,L \,,\qquad \frac{Q_{4}'(-\frac{i}{2})}{Q_{4}(-\frac{i}{2})}=\frac{3}{2}\,i\,L\, ,
\end{displaymath}
and thus verify that the anomalous dimensions \eqref{dimeng}
of the operators $\Tr \mathcal{F}^L$ are indeed, to one loop
\cite{Niklas thesis, super spin chain},
\begin{equation}
\Delta_{\mathcal{F}^L}=2\,L+6\,L\,g^2+\Op(g^4)\,.\nonumber
\end{equation}
Incidentally, 
it is straightforward to generalize the above method to all
higher conserved spin chain charges, by noting that
\begin{eqnarray}\label{conservedcharges}
q_r&=&\frac{i}{(r-1)(r-2)!}\left(\frac{\D^{r-1}}{\D u^{r-1}} \log Q_{4}(u)\arrowvert_{u=+\frac{i}{2}} - \frac{\D^{r-1}}{\D u^{r-1}} \log Q_{4}(u)\arrowvert_{u=-\frac{i}{2}}\right)\nonumber\\
&=&L\, \frac{i}{r-1}\left(1+\frac{1}{2^{r-1}}\right)(i^{r+1}-(-i)^{r+1})\,.
\end{eqnarray}
It is clear that all odd charges vanish. Note that the
$c_r$ in (3.20) of \cite{CompleteSpectrum} for the ``Beast'' diagram are given by $\frac{q_r}{L}$. 
\subsection{Generating Polynomials}
Starting from the set of all-loop Bethe ansatz equations (\ref{starting BE}) and dualizing the $u_3$ and $u_7$ roots 
\cite{LongRange} we obtain
\begin{eqnarray}
\left(\frac{x^{+}_{4,k}}{x^{-}_{4,k}}\right)^{L}&=&\prod^{2L-2}_{\substack{j=1\\j\neq k}} \frac{x^{-}_{4,k}-x^{+}_{4,j}}{x^{+}_{4,k}-x^{-}_{4,j}}\, \frac{1-\frac{g^2}{x^{+}_{4,k}x^{-}_{4,j}}}{1-\frac{g^2}{x^{-}_{4,k} x^{+}_{4,j}}}
\,\sigma^2(u_{4,k},u_{4,j})\, 
\prod^{2L-4}_{j=1} \frac{x^{+}_{4,k}-\tilde{x}_{5,j}}{x^{-}_{4,k}-\tilde{x}_{5,j}}\label{all1}\\
1&=&\prod^{L-2}_{j=1}\frac{\tilde{u}_{5,k}-u_{6,j}+\frac{i}{2}}{\tilde{u}_{5,k}-u_{6,j}-\frac{i}{2}} \prod^{2L-2}_{j=1}\frac{\tilde{x}_{5,k}-x^{+}_{4,j}}{\tilde{x}_{5,k}-x^{-}_{4,j}}\label{all2}\\
1&=&\prod^{L-2}_{\substack{j=1\\ j \neq k}} \frac{u_{6,k}-u_{6,j}+i}{u_{6,k}-u_{6,j}-i} \prod^{2L-4}_{j=1} \frac{u_{6,k}-\tilde{u}_{5,j}-\frac{i}{2}}{u_{6,k}-\tilde{u}_{5,j}+\frac{i}{2}}\label{all3}\, .
\end{eqnarray}
Here, $\tilde{u}_5$ are the $2L-4$ roots dual to $u_5$. 
Note that they are not given by (\ref{Q_5}) anymore.
It is however an easy task to derive the polynomials which generate 
the one-loop roots for this set of equations. One of them, generating 
the $u_4$ roots, is already known, see \eqref{Q}.
The polynomials generating the $\tilde{u}_{5}$ and the 
$u_{6}$ roots are given, respectively, by
\begin{eqnarray}\label{Q5}
Q_{5}(v)&=&3v^{2L}+(-i+v)^L (-2i+v)^L+(2i+v)^L 
\left((i+v)^L+(-2i+v)^L\right)\nonumber\\
&&-v^L \left((-i+v)^L+(i+v)^L+2(-2i+v)^L+2(2i+v)^L\right)
\end{eqnarray}
and
\begin{equation}\label{Q6}
Q_{6}(w)=(w+\tfrac{3}{2}i)^L+(w-\tfrac{3}{2}i)^L-(w+\tfrac{1}{2}i)^L-(w-\tfrac{1}{2}i)^L\, .
\end{equation}
Some properties of these polynomials are studied in \Appref{app:polyprop}.
%%%%%%%%%%%%%%%%%%%%%%%%%%%%%%%%%%%%%%%%%%%%%%%%%%%%%%%%%%%%%%%%%%%
\section{Thermodynamic Limit}
\label{sec:thermodynamic}

In this section we will demonstrate how to construct the thermodynamic 
limit and how to find the root densities using the generating polynomials defined in 
\eqref{Q},\eqref{Q5},\eqref{Q6}. 
We will present the method, taking $Q_{6}(w)$ in
\eqref{Q6} as an example.

Let us define the density of roots as
\begin{equation}\label{density definition}
\rho_{L,6} (z)=\frac{1}{L} \sum^{L-2}_{i=1} \delta(z-w_{i})\,.
\end{equation}
Using this definition we can write (\ref{Q6}) as
\begin{equation}
Q_{6}(w)=\prod^{L-2}_{i=1} (w-w_{i})=\exp\left(L \int^{\infty}_{-\infty} \rho_{L,6} (z)\log(w-z) dz \right)
\end{equation}
and easily obtain an integral equation for the density
\begin{equation}
\int^{\infty}_{-\infty} \frac{\rho_{L,6}(z)}{w-z}\,\D z=\frac{1}{L}\frac{Q_{6,L}(w)'}{Q_{6,L}(w)}=V_{L}(w)\,.
\end{equation}
Taking the limit $L \to \infty$ and using the identity
\begin{equation}
\frac{1}{x+i \epsilon}-\frac{1}{x-i\epsilon}=-2\pi i \delta(x)\,,
\end{equation}
we derive the formula
\begin{equation}\label{rhodurchv}
\rho_{6}(w)=\frac{i}{2 \pi} \left(V(w+i \epsilon)-V(w-i \epsilon)\right)\,,
\end{equation}
where 
\begin{equation}
\rho_{6}(u)=\lim_{L \to \infty} \rho_{L,6}(u)\qquad \text{and} \qquad V(w)=\lim_{L \to \infty} V_{L}(w)\,.
\end{equation}
It follows from the above formula that $V(w)$ is necessarily discontinuous 
across the real axis (i.e.~it has an infinite cut there).
To find $V(w)$ we compute
\begin{eqnarray}
V_{L}(w)&=&\frac{1}{L} \frac{Q_{6,L}'(w)}{Q_{6,L}(w)}\nonumber\\
&=&\frac{(w+\frac{3}{2}i)^{L-1}+(w-\frac{3}{2}i)^{L-1}-(w+\frac{1}{2}i)^{L-1}-(w-\frac{1}{2}i)^{L-1}}{(w+\frac{3}{2}i)^L+(w-\frac{3}{2}i)^L-(w+\frac{1}{2}i)^L-(w-\frac{1}{2}i)^L}\,.
\end{eqnarray}
Taking the limit is straightforward
\begin{equation}\label{V(w)}
V(w)\equiv \lim_{L \to \infty} V_{L} (w)= \left\{ \begin{array}{ll} 
\frac{1}{w+\frac{3}{2}i}\,, & \textrm{\scriptsize \;Im(w)$>$0}\\ 
\frac{1}{w-\frac{3}{2}i}\,, & \textrm{\scriptsize \;Im(w)$<$0}\end{array}\right.\,.
\end{equation}
Using (\ref{rhodurchv}) we find for the density
\begin{equation}\label{rho6}
\rho_{6}(w) \equiv \lim_{L \to \infty} \rho_{L,6} (w)=\frac{1}{2 \pi} \frac{3}{w^2+\frac{9}{4}}\,.
\end{equation}
The same procedure can be used to find the density for the momentum-carrying roots
\begin{equation}\label{rho4}
\rho_{4}(u)=\frac{1}{2 \pi} \left( \frac{1}{u^2+\frac{1}{4}}+\frac{3}{u^2+\frac{9}{4}} \right)\, ,
\end{equation}
which is a continuum limit of
\begin{equation}
\rho_{L,4} (z)=\frac{1}{L} \sum^{2L-2}_{i=1} \delta(z-u_{i})\, .
\end{equation}
As a validity check we can show that
\begin{eqnarray}
q_r&=&L \frac{i}{r-1} \int^{\infty}_{-\infty} \rho_{4} (u) \left(\frac{1}{(u+\frac{i}{2})^{r-1}}-\frac{1}{(u-\frac{i}{2})^{r-1}} \right)\nonumber\\
&=&L \frac{i}{r-1}\left(1+\frac{1}{2^{r-1}}\right)(i^{r+1}-(-i)^{r+1})\,,
\end{eqnarray}
which is in agreement with the earlier result (\ref{conservedcharges}).

Let us now direct our attention to the density of roots of flavor $u_5$. {}From the analysis of the generating polynomial we infer that these roots form two strings along $\pm\frac{i}{2}$ (see \Appref{app:stacks} for details) on the complex plane and we have to integrate over two contours
\begin{equation}\label{contour}
\int^{\infty +\frac{i}{2}}_{-\infty+\frac{i}{2}} \frac{\rho_{5}(v^{+})}{u-v^{+}}\D v^{+}+\int^{\infty-\frac{i}{2}}_{-\infty-\frac{i}{2}} \frac{\rho_{5}(v^{-})}{u-v^{-}}\D v^{-}=\Lambda(u)\,,
\end{equation}
where $\Lambda(u)$ is given by
\begin{equation}
\Lambda(u)=\lim_{L \to \infty}\tfrac{1}{L}~\partial_{u}\log{Q_{5,L}(u)}=
\left\{ \begin{array}{ll} 
\frac{1}{u+i}+\frac{1}{u+2i}\,, & \textrm{\scriptsize \;Im(u)$>\frac{1}{2}$}\\ 
\frac{1}{u-2i}+\frac{1}{u+2i}\,, & \textrm{\scriptsize \;$-\frac{1}{2}<$ Im(u)$<\frac{1}{2}$}\\
\frac{1}{u-i}+\frac{1}{u-2i}\,, & \textrm{\scriptsize \;Im(u)$<-\frac{1}{2}$}
\end{array}\right.\,.
\end{equation}
The solution of (\ref{contour}) is
\begin{equation}\label{stack}
\rho_{5}(v\pm\tfrac{i}{2})=\rho_{6}(v)\,.
\end{equation}
Hence, two roots $u_{5}^{\pm}$ form a {\it stack} with one root of type $u_{6}$ in the center such that \mbox{$u_{5}^{\pm}=(u_{6} \pm \tfrac{i}{2})$}. The concept of stacks was introduced and studied in
the context of the nested Bethe ansatz of 
``ferromagnetic'' root distributions \cite{CompleteSpectrum}.
Here we find the same stack picture in the context of a long operator
akin to an ``antiferromagnetic'' state.
The emergence of the stack picture for our case
is rigorously established in \Appref{app:stacks}.
%
%%%%%%%%%%%%%%%%%%%%%%%%%%%%%%%%%%%%%%%%%%%%%%%%%%%%%%%%%%%%%%%
%
\subsection{Asymptotic All-Loop Effective Bethe Equations}
\
In the thermodynamic limit only the centers of the stacks receive quantum corrections, 
i.e.~\mbox{$u_{5,k}(g)\approx u_{6,k}(g)+\tfrac{i}{2}$} and 
\mbox{$u_{5,k+L-2}(g)\approx u_{6,k}(g)-i/2$}
for \mbox{$k=1, ..., L-2$}. The effective set of Bethe equations then reads\footnote{The use of "$\approx$" instead of "$=$" 
indicates that these equations are strictly speaking not exactly valid 
for finite $L$ due to differences of $\mathcal{O}(\frac{1}{L})$ for the generating polynomials. 
See \Appref{app:stacks} for more details.}
\begin{eqnarray}
\left(\frac{x^{+}_{4,k}}{x^{-}_{4,k}}\right)^{L}&\approx&\prod^{2L-2}_{\substack{j=1\\j\neq k}} \frac{x^{-}_{4,k}-x^{+}_{4,j}}{x^{+}_{4,k}-x^{-}_{4,j}}\, 
\frac{1-g^{2}/x^{+}_{4,k}x^{-}_{4,j}}{1-g^{2}/x^{-}_{4,k} x^{+}_{4,j}}\,
\sigma^2(u_{4,k},u_{4,j})\,
\nonumber\\
&&\times
\prod^{L-2}_{j=1} \frac{x^{+}_{4,k}-x^{-}_{6,j}}{x^{-}_{4,k}-x^{+}_{6,j}}
\frac{1-g^{2}/x^{-}_{4,k}x^{-}_{6,j}}{1-g^{2}/x^{+}_{4,k} x^{+}_{6,j}}\label{alleffective1}
\end{eqnarray}
\begin{equation}\label{alleffective2}
1\approx \prod^{L-2}_{\substack{j=1\\j \neq k}} \frac{u_{6,k}-u_{6,j}+i}{u_{6,k}-u_{6,j}-i} \prod^{2L-2}_{j=1}\frac{x^{-}_{6,k}-x^{+}_{4,j}}{x^{+}_{6,k}-x^{-}_{4,j}} \frac{1-g^{2}/x^{-}_{6,k}x^{-}_{4,j}}{1-g^{2}/x^{+}_{6,k} x^{+}_{4,j}}\, ,
\end{equation}
where we have used the identity
\begin{displaymath}
(x^{\pm}_{k}-x^{\pm}_{j})=\frac{u_{k}-u_{j}}{1-g^{2}/x^{\pm}_{k}x^{\pm}_{j}}\,.
\end{displaymath}
Note that (\ref{all2}) splits up into two equations with $k$ running from $k=1,\dots,L-2$ and $k=L-2+1,\ldots,2L-2$ respectively, which can by multiplied with each other to get (\ref{alleffective2}), while equation (\ref{all3}) is identically satisfied. 
These effective equations describe the scattering of the roots $u_4$ with the centers of the stacks.\
Rewriting (\ref{alleffective1}) and (\ref{alleffective2}) as 
\begin{eqnarray}\label{transform1}
\left(\frac{x^{+}_{4,k}}{x^{-}_{4,k}}\right)^{L}&\approx&\prod^{2L-2}_{\substack{j=1\\j\neq k}} \frac{u_{4,k}-u_{4,j}-i}{u_{4,k}-u_{4,j}+i}\,
\left(\frac{1-g^{2}/x^{+}_{4,k}x^{-}_{4,j}}{1-g^{2}/x^{-}_{4,k}x^{+}_{4,j}}\right)^2\,\sigma^2(u_{4,k},u_{4,j})
\nonumber\\
&&\times \prod^{L-2}_{j=1}\frac{u_{4,k}-u_{6,j}+i}{u_{4,k}-u_{6,j}-i}\,
\frac{1-g^{2}/x^{-}_{4,k}x^{+}_{6,j}}{1-g^{2}/x^{+}_{4,k}x^{-}_{6,j}}\,
\frac{1-g^{2}/x^{-}_{4,k}x^{-}_{6,j}}{1-g^{2}/x^{+}_{4,k} x^{+}_{6,j}}
\end{eqnarray}
\begin{eqnarray}\label{transform2}
1\approx&&\prod^{L-2}_{\substack{j=1\\j\neq k}} \frac{u_{6,k}-u_{6,j}+i}{u_{6,k}-u_{6,j}-i} 
\nonumber\\
&&\times
\prod^{2L-2}_{j=1}\frac{u_{6,k}-u_{4,j}-i}{u_{6,k}-u_{4,j}+i},
\frac{1-g^{2}/x^{+}_{6,k}x^{-}_{4,j}}{1-g^{2}/x^{-}_{6,k} x^{+}_{4,j}}\, 
\frac{1-g^{2}/x^{-}_{6,k}x^{-}_{4,j}}{1-g^{2}/x^{+}_{6,k} x^{+}_{4,j}}\,.
\end{eqnarray}
and taking the logarithm of these equations leads to
\begin{eqnarray}\label{log1}
&&\frac{1}{i}\left (\log x^{+}_{4,k}-\log x^{-}_{4,k} \right)\approx 2\pi\frac{n_{k}}{L}
+\frac{1}{iL}\sum^{2L-2}_{\substack{j=1\\j\neq k}} 
\log\frac{u_{4,k}-u_{4,j}-i}{u_{4,k}-u_{4,j}+i}
\nonumber\\
&&
+\frac{1}{iL}\sum^{2L-2}_{\substack{j=1\\j\neq k}} \left(
2\, \log\frac{1-g^{2}/ x^{+}_{4,k}x^{-}_{4,j}}{1-g^{2}/x^{-}_{4,k}x^{+}_{4,j}}+2\,i\, \theta(u_{4,k},u_{4,j})\right)
\nonumber\\
&&+\frac{1}{iL}\sum^{L-2}_{j=1}\left(
\log \frac{u_{4,k}-u_{6,j}+i}{u_{4,k}-u_{6,j}-i}
+\log \frac{1-g^{2}/x^{-}_{4,k}x^{+}_{6,j}}{1-g^{2}/x^{+}_{4,k}x^{-}_{6,j}}+
\log\frac{1-g^{2}/x^{-}_{4,k}x^{-}_{6,j}}{1-g^{2}/x^{+}_{4,k} x^{+}_{6,j}}
\right)\, ,
\end{eqnarray}
where the dressing phase shift $2\,i\,\theta$ is given in \eqref{dressingfactor}, and to
\begin{eqnarray}\label{log2}
0&\approx&2\pi\frac{m_{k}}{L}+\frac{1}{iL}\sum^{L-2}_{\substack{j=1\\j\neq k}} \log\frac{u_{6,k}-u_{6,j}+i}{u_{6,k}-u_{6,j}-i}\nonumber\\
&+&\frac{1}{iL}\sum^{2L-2}_{j=1}\left(
\log\frac{u_{6,k}-u_{4,j}-i}{u_{6,k}-u_{4,j}+i}+
\log\frac{1-g^{2}/x^{+}_{6,k}x^{-}_{4,j}}{1-g^{2}/x^{-}_{6,k} x^{+}_{4,j}}+\log\frac{1-g^{2}/x^{-}_{6,k}x^{-}_{4,j}}{1-g^{2}/x^{+}_{6,k} x^{+}_{4,j}}
\right)\,.
\end{eqnarray}
The ambiguity in the choice of branch of the logarithm is encoded in the mode numbers $n_k$ and $m_k$, for which we introduce mode functions $n_{k}=n(u_{4,k})$ and $m_{k}=m(u_{6,k})$.\\
As usual we will now replace the sums by integrals as $L \to \infty$ and introduce new variables $\xi=\frac{n(u_{4})}{L}$ and $\chi=\frac{m(u_{6})}{L}$. They allow us to define the all-loop excitation 
densities through
$\rho(u)=-\frac{\D\xi(u)}{\D u}$ and $\eta(w)=-\frac{\D\chi(w)}{\D w}$. One can show that this is consistent as the correct one-loop densities can be derived from the effective equations. Finally we take the derivatives w.r.t.~$u$ and $w$, respectively.  
Thus, (\ref{log1}) and (\ref{log2}) become:
\begin{eqnarray}\label{alllooptosolve1}
&&\frac{1}{i}\left (\frac{1}{\sqrt{u^{2}_{-}-4g^{2}}}-\frac{1}{\sqrt{u^{2}_{+}-4g^{2}}}\right)=2\,\pi\, \rho(u)
-2\int^{\infty}_{-\infty}\frac{\rho(u')\D u'}{(u-u')^2+1}\nonumber\\
&&+2i\int^{\infty}_{-\infty}\D u'~\rho(u')\frac{\D}{\D u}\left(
\log\frac{1-g^{2}/x^{+}(u)x^{-}(u')}{1-g^{2}/x^{-}(u)x^{+}(u')}+
i\,\theta(u,u')\right)
\nonumber\\
&&+2\int^{\infty}_{-\infty}\frac{\eta(w)\D w}{(u-w)^2+1}\nonumber\\
&&+i\int^{\infty}_{-\infty}\D w~\eta(w)\frac{\D}{\D u}\left(
\log\frac{1-g^{2}/x^{-}(u)x^{+}(w)}{1-g^{2}/x^{+}(u)x^{-}(w)}+
\log\frac{1-g^{2}/x^{-}(u)x^{-}(w)}{1-g^{2}/x^{+}(u)x^{+}(w)}\right)
\end{eqnarray}
and
\begin{eqnarray}\label{alllooptosolve2}
0&=&2\,\pi\, \eta(w)+2\int^{\infty}_{-\infty}\frac{\eta(w')\D w'}{(w-w')^2+1}-2\int^{\infty}_{-\infty}\frac{\rho(u)\D u}{(w-u)^2+1}\nonumber\\
&&+i\int^{\infty}_{-\infty}\D u~\rho(u)\frac{\D}{\D w}\left(
\log\frac{1-g^{2}/x^{+}(w)x^{-}(u)}{1-g^{2}/x^{-}(w)x^{+}(u)}+
\log\frac{1-g^{2}/x^{-}(w)x^{-}(u)}{1-g^{2}/x^{+}(w)x^{+}(u)}\right)\,.
\end{eqnarray}
%
%%%%%%%%%%%%%%%%%%%%%%%%%%%%%%%%%%%%%%%%%%%%%%%%%%%%%%%%%%%%%%%%%%%%
%
\subsection{All-Loop Energy}
Our strategy is now to solve \eqref{alllooptosolve2}
for the auxiliary root density $\eta$ 
as a functional of the energy-momentum-carrying main root density 
$\rho$. Substitution of this solution into \eqref{alllooptosolve2} 
then yields a single, closed equation for the latter.
This is easily done by Fourier transformation
techniques, see in particular \cite{InteTrans}. 

The Fourier transforms of the densities $\rho(u)$ and $\eta(w)$
are defined as
\begin{equation}\label{denstrans}
\density(t)=
e^{-\frac{|t|}{2}}\,\int^{\infty}_{-\infty}e^{i\, t\,u}\,\rho(u)\D u \, ,
\qquad
\auxdensity(t)=
e^{-\frac{|t|}{2}}\,\int^{\infty}_{-\infty}e^{i\, t\,w}\,\eta(w)\D w \, ,
\end{equation}
where we have included a factor $e^{-|t|/2}$ for convenience.
Including the same prefactors, the double-Fourier transforms of the kernels
\begin{eqnarray}
K(w,u)=\frac{1}{2 \pi i}~\partial_{w}\log\frac{1-g^{2}/x^{+}(w)x^{-}(u)}{1-g^{2}/x^{-}(w)x^{+}(u)}\label{SKernel}\,,\\
H(w,u)=\frac{1}{2 \pi i}~\partial_{w}\log\frac{1-g^{2}/x^{-}(w)x^{-}(u)}{1-g^{2}/x^{+}(w)x^{+}(u)}\label{ASKernel}\,,
\end{eqnarray}
may be explicitly computed as, respectively,
\begin{eqnarray}
\hat{K}(t,t')=&&2\pi g^{2}(1-\sign~tt')|t|e^{-|t|-|t'|}
\frac{\BesselJ_{0}(2g|t|)\BesselJ_{1}(2g|t'|)-\BesselJ_{0}(2g|t'|)\BesselJ_{1}(2g|t|)}{2g(|t|-|t'|)}\label{Khat}\,,\\
\hat{H}(t,t')=&-&2\pi g^{2}(1+\sign~tt')|t|e^{-|t|-|t'|}
\frac{\BesselJ_{0}(2g|t|)\BesselJ_{1}(2g|t'|)+\BesselJ_{0}(2g|t'|)\BesselJ_{1}(2g|t|)}{2g(|t|+|t'|)}\label{Hhat}\,.
\end{eqnarray}
Fourier transforming \eqref{alllooptosolve2} diagonalizes
all terms containing $\eta$, and one obtains
\begin{eqnarray}\label{alllooptoone2fourier}
0&=&e^{|t|}\,\auxdensity(t)+\auxdensity(t) -\density(t)\nonumber\\
&&-\frac{1}{2\pi}\int^{\infty}_{-\infty}\D t'\,
\density(-t')\,e^{|t|}\,\hat{K}(t,t')\,e^{|t'|}-\frac{1}{2\pi}\int^{\infty}_{-\infty}\D t'\,
\density(-t')\,e^{|t|}\,\hat{H}(t,t')\,e^{|t'|}\,.
\end{eqnarray}
As announced above, we may now solve for $\auxdensity(t)$ in 
terms of an integral transform of $\density(t)$
\begin{equation}\label{inteta}
\auxdensity(t)=\frac{1}{e^t+1}\left(\density(t)-
4\,g^{2}\,t\int^{\infty}_{0}\D t' \,\hat{K}_{0}(2gt,2gt')\,\density(t')\right )\,,
\end{equation}
whose kernel $\hat{K}_{0}(t,t')$ is defined as in \eqref{K0}.

Now we treat (\ref{alllooptosolve1}) in a likewise fashion. The Fourier transform of the l.h.s.~is given by
\begin{equation}
\frac{1}{2\pi i}\int_{-\infty}^{\infty}\D u\left (\frac{1}{\sqrt{u^{2}_{-}-4g^{2}}}-\frac{1}{\sqrt{u^{2}_{+}-4g^{2}}}\right)e^{itu}=e^{-|t|/2}\BesselJ_{0}(2gt).
\end{equation}
Thus, equation (\ref{alllooptosolve1}) reads in Fourier space 
\begin{eqnarray}\label{almostthere}
\BesselJ_{0}(2gt)&=&e^{t}\,\density(t)-\density(t)+\auxdensity(t)
\nonumber\\
&&+4\,g^{2}\,t \int^{\infty}_{0}\D t'\, 
\left(\hat{K}_{m}(2gt,2gt')+\hat{K}_{d}(2gt,2gt')\right)
\density(t')\nonumber\\
&&-4\,g^{2}\,t\int^{\infty}_{0}\D t'\,\hat{K}_{1}(2gt,2gt')\,\auxdensity(t')\,,
\end{eqnarray}
where we have used \eqref{dressingfactor}, 
and $\hat{K}_{1}$ is given in \eqref{K1}
while $\hat{K}_{m}=\hat{K}_0+\hat{K}_1$ is
\begin{equation}\label{Km}
\hat{K}_{m}(t,t')=\frac{\BesselJ_{1}(t)\,\BesselJ_{0}(t')-
\BesselJ_{0}(t)\,\BesselJ_{1}(t')}{t-t'}\, .
\end{equation}
Now we may substitute \eqref{inteta} into \eqref{almostthere}.
This leads to a single equation for $\density(t)$
\begin{eqnarray}\label{integralequation}
\density(t)&=&e^{-2t}(1+e^{t})\bigg ( \BesselJ_{0}(2gt)
\nonumber\\
&&+4g^{2}t\int^{\infty}_{0}\D t'\Big ( \frac{1}{e^{t}+1}\hat{K}_{0}(2 g t,2 g t')+\frac{1}{e^{t'}+1}\hat{K}_{1}(2 g t,2 g t')\Big)\density(t')
\nonumber\\
&&-4g^{2}t\int^{\infty}_{0}\D t'\left (\hat{K}_{m}(2 g t,2 g t')+\hat{K}_n(2 g t,2 g t')+\hat{K}_{d}(2gt,2gt')\right)\density(t')\bigg)\,,\nonumber\\
\end{eqnarray}
as announced in \eqref{finaleq} at the beginning of this paper.
Since we have two degrees of freedom per
unit length, i.e.~the state is doubly-filled,
the density \eqref{integralequation}
is correspondingly normalized to $2$, 
i.e.~$\density(0)=2$, see \eqref{denstrans}.
The dressing kernel $\hat{K}_{d}(t,t')$ is written in \eqref{Kd},
and we just proved that the nesting kernel
$\hat{K}_n(t,t')$ is indeed given by the structurally very
similar expression \eqref{Kn}.
In fact,  $2\hat{K}_n$ and $\hat{K}_{d}$ differ only by a single sign in the denominator of the diagonal kernel $t/(e^t \pm1)$. 

The energy of the state can now be computed from the
density by Fourier transforming \eqref{chainenergy}. It reads
\begin{equation}\label{energy}
E(g)=4L\int^{\infty}_{0}\D t~\frac{\BesselJ_{1}(2gt)}{2gt}~\density(t)\,.
\end{equation}
This leads via \eqref{dimeng} to the anomalous dimension \eqref{dim} announced at the beginning.

We expect our solution to be valid at {\it arbitrary} values
of the coupling constant. The integral equation 
\eqref{integralequation} is however
(presumably) too complicated to be explicitly solvable.
Being of Fredholm-type, it is however very easy to
expand in small $g$ to high orders.  
We thus find for the perturbative
energy to e.g.~seven-loop order
\begin{eqnarray}\label{dimexp}
\frac{E(g)}{L}&=&3-\frac{51}{4} g^{2}+\frac{393}{4}\,g^{4}-\left(\frac{59487}{64}+54\,\zeta(3)\right) g^{6}\nonumber\\
&&+\left(\frac{632661}{64}+\frac{1665}{2}\,\zeta(3)+540\,\zeta(5)\right)g^{8}\nonumber\\
&&-\left(\frac{29056443}{256}+\frac{87525}{8}\,\zeta(3)+8505\,\zeta(5)+5670\,\zeta(7)\right)g^{10}+\nonumber\\
&&+\bigg(\frac{351914817}{256}+\frac{2244573}{16}\,\zeta(3)+972\,\zeta(3)^{2}\nonumber\\
&&\ \ \ \ +114723\,\zeta(5)+90909\,\zeta(7)+63504\,\zeta(9)\bigg)g^{12}+\ldots\, .
\end{eqnarray}
We notice that zeta functions of odd, but not even, argument
enter the energy.
This is not surprising, as these are directly generated by
the dressing kernel \eqref{dressingfactor}, {\it cf} \cite{TransCross}.
If we assign a ``degree of transcendentality'' $k$ to $\zeta(k)$
we see that the contributions at a given loop order are,
in contradistinction to the case of large-spin twist operators,
see \cite{InteTrans,TransCross,highest trans}, 
{\it not} of constant degree.
One may nevertheless observe that at a given order $l$ the
degree is bounded, and always saturated, by $2l-5$.
Note also that all zeta-function coefficients, as well
as all rational numbers, turn out to be integers after 
factoring out inverse powers of 2.

Conversely, the dressing factor is the {\it only} source
of $\zeta$-function terms in the expansion \eqref{dimexp}.
By this we mean that dropping the dressing factor from
the asymptotic Bethe ansatz equations \eqref{starting BE}
would eradicate all terms containing $\zeta$-functions
in \eqref{dimexp}, and would generate only rational 
loop contributions. The latter would precisely agree
with all terms of ``transcendentality degree zero'' in
\eqref{dimexp}. 

It is also interesting to note that the thermodynamic expansion of 
\eqref{dimexp} does not coincide with the energies of finite length 
operators (see \Appref{app:numres}), as opposed to the
case of the pseudo-vacuum state $\Tr \fldU^L$ 
\cite{HES SU(1|1),Dual Bethe SU(1|1)}.
By this we mean that the exact anomalous dimension of
the operator $\Tr \fldF^L$, even below wrapping order,
is not exactly proportional to $L$, while it is for $\Tr \fldU^L$.
This is very likely due to the length changing processes starting 
for field strength operators at two-loop order, 
i.e.~the number of fields in a local operator is not a conserved quantity 
at higher loops. Correspondingly, and in contradistinction to
a BPS-state $\Tr \fldZ^L$ or a fermionic 
pseudo-vacuum $\Tr \fldU^L$, the one-loop field strength
operator $\Tr \fldF^L$ is not an exact eigenstate  and will pick up 
higher-order quantum corrections: $\Tr \mathcal{F}^L + {\cal O}(g^2)$.

%%%%%%%%%%%%%%%%%%%%%%%%%%%%%%%%%%%%%%%%%%%%%%%%%%%%%%%%%%%%%%%%%%
\section{Maximal Filling and Nesting}
\label{sec:SO6singlet}

We would like to demonstrate that the emergence of a nesting
kernel \eqref{Kn}, which we argue to be closely 
analogous to a dressing kernel \eqref{Kd}, is a rather generic mechanism if {\it two} prerequisites
are met. The first is that the states satisfy a special maximal filling 
condition, and the second is that the state is irreducibly\footnote{
By this we mean that the nesting cannot be removed by an
exact dualization of the Bethe roots.} nested.

Let us first study two simple cases where only the first, but not
the second prerequisite is fulfilled. These are the ``highest energy"
antiferromagnetic state of the $\mathfrak{su}(2)$ sector
\cite{hubbard,zarembo su2},
and the fermionic pseudo-vacuum state $\Tr \fldU^L$
\cite{HES SU(1|1),Dual Bethe SU(1|1)}. As a by-product we
will find the correct perturbative expansion of these states,
as the expressions in \cite{hubbard,zarembo su2,HES SU(1|1)}
were obtained with a trivial dressing factor $\sigma^2=1$,
and need to be revised starting at four loops.

For the $\mathfrak{su}(2)$ antiferromagnet the occupation
numbers for an even length $L$ operator are
\begin{equation}
(K_1,\,K_2,\, K_3, \,K_4, \,K_5, \,K_6, \,K_7)=
(0,\,0,\, 0,\,\frac{L}{2},\,0,\,0,\, 0).
\end{equation}
It is straightforward to repeat the analysis
of \cite{hubbard,zarembo su2} in the presence of the
non-trivial dressing phase \eqref{dressingfactor}, and one finds
that the Fourier-transformed density of roots satisfies in the 
thermodynamic limit the linear integral equation
\begin{equation}\label{su2equation}
\density(t)=\frac{1}{e^t+1}\left( \BesselJ_{0}(2gt)-4\,g^{2}\,t
\int_{0}^{\infty}\D t'\,\hat{K}_{d}(2gt,2gt')\,\density(t') \right)\, .
\end{equation}
This is a half-filled state, which is the maximally possible
filling in this sector. Correspondingly, the density is normalized
to $1/2$, i.e.~$\density(0)=\tfrac{1}{2}$, see \eqref{denstrans}.
The energy is as always given by \eqref{energy}. 
Dropping the convolution term on the r.h.s.~of \eqref{su2equation}
yields the density for the Lieb-Wu ground-state energy of the 
fermionic Hubbard
model in a closed form, see \cite{hubbard}. 
The $\alg{su}(2)$ sector of $\superN=4$ gauge theory
is a supersymmetric deformation of the latter. It is described by adding the
backreacting convolution. Apparently, the density 
$\density(t)$ can no longer
be found in closed form. However, it is interesting to
work out the weak coupling expansion of our model from
\eqref{energy},\eqref{su2equation}. One finds to the first few orders
\begin{eqnarray}\label{su2eng}
\frac{E(g)}{L}&=&2\, \zeta_{a}(1) - 6\,\zeta_{a}(3)g^2 + 40\, \zeta_{a}(5)\,g^4 -\bigg(350\, \zeta_{a}(7) + 32\, \zeta_{a}(1)\, \zeta_{a}(3)\, \zeta(3)\bigg)g^6 
\nonumber\\
&&
  + \bigg(3528\, \zeta_{a}(9) + 96\, \zeta_{a}(3)^2 \, \zeta(3) + 320\, \zeta_{a}(1)\, \zeta_{a}(5)\, \zeta(3) + 320\, \zeta_{a}(1)\, \zeta_{a}(3)\, \zeta(5)\bigg)g^8 \nonumber\\
&& -\bigg(38808\, \zeta_{a}(11) + 1600\, \zeta_{a}(3)\, \zeta_{a}(5)\, \zeta(3) + 3360\, \zeta_{a}(1)\, \zeta_{a}(7)\, \zeta(3) 
\nonumber\\
&&\ \ \ \ + 1024 \, \zeta_{a}(3)^2 \,\zeta(5)+3264\, \zeta_{a}(1)\, \zeta_{a}(5)\, \zeta(5) + 3360\, \zeta_{a}(1)\, \zeta_{a}(3)\, \zeta(7))\bigg)g^{10}\nonumber\\
&&+  \bigg(453024\, \zeta_{a}(13) + 6400 \, \zeta_{a}(5)^2 \, \zeta(3)+ 15680\, \zeta_{a}(3)\, \zeta_{a}(7)\, \zeta(3)
\nonumber\\
&&\ \ \ \ + 37632 \, \zeta_{a}(1)\, \zeta_{a}(9)\, \zeta(3) +  512\, \zeta_{a}(1)\, \zeta_{a}(3)^2 \,\zeta(3)^2 
+ 17792\, \zeta_{a}(3)\, \zeta_{a}(5)\, \zeta(5) \nonumber\\
&&\ \ \ \ + 34944\, \zeta_{a}(1)\, \zeta_{a}(7)\, \zeta(5) + 11200\, \zeta_{a}(3)^2 \,\zeta(7) + 34944\, \zeta_{a}(1)\, \zeta_{a}(5)\, \zeta(7) \nonumber\\
&&
\ \ \ \ + 37632\, \zeta_{a}(1)\, \zeta_{a}(3)\, \zeta(9)\bigg)g^{12}+\ldots\,.
\end{eqnarray}
Here, we have decided to distinguish the alternating 
``fermionic'' $\zeta_a$-function
\begin{equation}\label{zetaaltdef}
\zeta_{a}(k)=\frac{1}{\Gamma(k)}\int_0^\infty \frac{dt}{t}\,\frac{t^k}{e^t+1}=
\sum_{n=1}^{\infty}\frac{(-1)^{n-1}}{n^k}\,.
\end{equation}
from the ordinary ``bosonic'' $\zeta$-function
\begin{equation}\label{zetadef}
\zeta(k)=\frac{1}{\Gamma(k)}\int_0^\infty \frac{dt}{t}\,\frac{t^k}{e^t-1}=
\sum_{n=1}^{\infty}\frac{1}{n^k}\,,
\end{equation}
even though they are related by the formula 
$\zeta_a(k)=(1-2^{1-k})\,\zeta(k)$. Note also that $\zeta_a(1)=\log(2)$.
Using these relations we can ``simplify'' the expression \eqref{su2eng} to
\begin{eqnarray}\label{su2engalt}
\frac{E(g)}{L}&=&2\, \log(2)-\frac{9}{2}\, \zeta(3)g^2+\frac{75}{2}\, \zeta(5)g^4-\bigg(24\, \log(2)\,\zeta(3)^2+\frac{11025}{32}\, \zeta(7) \bigg)g^6\nonumber\\
&&+\bigg(54\, \zeta(3)^3+540\, \log(2)\, \zeta(3)\, \zeta(5)+\frac{112455}{32}\, \zeta(9) \bigg)g^8\nonumber\\
&&-\bigg(1701\, \zeta(3)^2\, \zeta(5)+3060\, \log(2)\,\zeta(5)^2+\frac{11655}{2}\, \log(2)\, \zeta(3)\, \zeta(7)\nonumber\\
&&\ \ \ \ +\frac{4962573}{128}\, \zeta(11) \bigg)g^{10}\nonumber\\
&&+\bigg(288\, \log(2)\, \zeta(3)^4+18135\, \zeta(3)\, \zeta(5)^2+\frac{71505}{4}\,\zeta(3)^2 \, \zeta(7)+67158\, \log(2)\, \zeta(5)\, \zeta(7)\nonumber\\
&&\ \ \ \  +65709\, \log(2)\, \zeta(3)\, \zeta(9)+\frac{57972915}{128}\, \zeta(13) \bigg)g^{12}+\ldots\,.
\end{eqnarray}
This however obscures the distinction between the contributions 
stemming from the dressing factor (the $\zeta$-terms) and from 
the fermionic Hubbard model (the  $\zeta_a$-terms). 
In fact, omitting all terms containing
$\zeta$ (in \eqref{su2eng}, but not in \eqref{su2engalt}!)
leads back to the Hubbard ground-state energy.
This admixture of bosonic $\zeta$ and fermionic 
$\zeta_a$ is further evidence
that planar AdS/CFT is a supersymmetric generalization of
the purely fermionic Hubbard model as employed in \cite{hubbard}.
Note also that the coefficients multiplying the $\zeta$- and
$\zeta_a$-functions in \eqref{su2eng} are all integers. 
Interestingly, we see that the terms in \eqref{su2eng}
are still of constant degree of
transcendentality $(2l-1)$ at a given loop order $l$
if we assign a ``degree of transcendentality'' $k$ to both
$\zeta(k)$ and $\zeta_a(k)$. 

We now turn to the $\alg{su}(1|1)$ state $\Tr \fldU^L$.
The occupation numbers are 
\begin{equation}
(K_1,\,K_2,\, K_3, \,K_4, \,K_5, \,K_6, \,K_7)=
(0,\,0,\, 0,\,L,\,L-1,\,0,\, 0).
\end{equation}
In the picture of Fig.~1, we first replace all $L$ fields $\fldZ$
in the BPS vacuum $\Tr \fldZ^L$ by $L$ bosons $\fldX$,
and then turn the bosons into fermions $\fldU$.
This state may be dualized, and the two-level nested Bethe equations 
may be converted to a single level \cite{LongRange}. 
Extending the analysis of \cite{HES SU(1|1)} to the case of a
non-trivial dressing phase \eqref{dressingfactor}, we find
that the Fourier-transformed density of roots satisfies in the 
limit $L \rightarrow \infty$ the equation
\begin{equation}\label{su11equation}
\density(t)=e^{-t}\,\left (\BesselJ_{0}(2gt)-2\,g^{2}\,t
\int_{0}^{\infty}\D t'\, \big(\hat{K}_{m}(2gt,2gt')
+2\hat{K}_{d}(2gt,2gt')\big)\density(t') \right)\,.
\end{equation}
This is a filled state, with one excitation per lattice site, 
which is the maximally possible filling in this sector. 
Correspondingly, the density is normalized
to $1$, i.e.~$\density(0)=1$.
Working out the weak coupling expansion of this state from
\eqref{energy},\eqref{su11equation}, 
one finds to e.g.~eight-loop order
\begin{eqnarray}
\frac{E(g)}{L}&=&2-8g^2+58g^4-\bigg(518+32\, \zeta(3)\bigg)g^6+\bigg(5228+480\, \zeta(3)+320\, \zeta(5)\bigg)g^8\nonumber\\
&&-\bigg(57280+6144\, \zeta(3)+4928\, \zeta(5)+3360\, \zeta(7)\bigg)g^{10}\nonumber\\ &&+\bigg(665344+76768\, \zeta(3)+512\, \zeta(3)^2+64960\, \zeta(5)\nonumber\\
&&\ \ \ \ +52864\, \zeta(7)+37632\, \zeta(9)\bigg)g^{12}\nonumber\\
&&-\bigg(8070352+965856\, \zeta(3)+13312\, \zeta(3)^2+833792\, \zeta(5)+10240\, \zeta(3)\, \zeta(5)\nonumber\\
&&\ \ \ \ +713056\, \zeta(7)+602112\, \zeta(9)+443520\, \zeta(11)\bigg)g^{14}+\ldots\,.
\end{eqnarray}
The $\zeta$-functions are exclusively generated by the dressing factor.
We note similar ``transcendentality properties'' as in the 
$\Tr \fldF^L$ case of the last section, namely
that at a given loop order $l$ combinations of zeta functions with odd 
arguments occur up to and including degree of transcendentality
$2l-5$. Again all zeta-function coefficients, as well
as all rational numbers, turn out to be integers.

Let us finally present a third example, 
namely a certain $\mathfrak{so}(6)$ singlet state. At one-loop this state 
is the highest energy state of a $\mathfrak{so}(6)$ magnet 
\cite{Minahan&Zarembo,Resh}. 
At higher loops, however, the $\mathfrak{so}(6)$ subsector is not closed 
anymore and thus one is forced to use the full $\mathfrak{psu}(2,2|4)$ Bethe equations.  
The excitation scheme for this state reads 
\begin{equation}
(K_1,\,K_2,\, K_3, \,K_4, \,K_5, \,K_6, \,K_7)=(\frac{L}{2}-2,\,\frac{L}{2}-1,\,\frac{L}{2},\,L,\,\frac{L}{2},\,\frac{L}{2}-1,\,\frac{L}{2}-2).
\end{equation}
After the dualization of the $u_5$ and $u_3$ roots one is left with the following equations
\begin{eqnarray}\label{eq:SO6}
1&=&\prod^{\frac{L}{2}-1}_{\substack{j=1\\ j \neq k}} \frac{u_{2,k}-u_{2,j}+i}{u_{2,k}-u_{2,j}-i} \prod^{L-2}_{j=1} \frac{u_{2,k}-\tilde{u}_{3,j}-\frac{i}{2}}{u_{2,k}-\tilde{u}_{3,j}+\frac{i}{2}}\nonumber\\
1&=&\prod^{\frac{L}{2}-1}_{j=1}\frac{\tilde{u}_{3,k}-u_{2,j}+\frac{i}{2}}{\tilde{u}_{3,k}-u_{2,j}-\frac{i}{2}} \prod^{L}_{j=1}\frac{\tilde{x}_{3,k}-x^{+}_{4,j}}{\tilde{x}_{3,k}-x^{-}_{4,j}}\nonumber\\
\left(\frac{x^{+}_{4,k}}{x^{-}_{4,k}}\right)^{L}&=&
\prod^{L}_{\substack{j=1\\j\neq k}} \frac{x^{-}_{4,k}-x^{+}_{4,j}}{x^{+}_{4,k}-x^{-}_{4,j}} \frac{1-\frac{g^2}{x^{+}_{4,k}x^{-}_{4,j}}}{1-\frac{g^2}{x^{-}_{4,k} x^{+}_{4,j}}}\,\sigma^2(u_{4,k},u_{4,j}) 
\prod^{L-2}_{j=1} \frac{x^{+}_{4,k}-\tilde{x}_{3,j}}{x^{-}_{4,k}-\tilde{x}_{3,j}}\prod^{L-2}_{j=1} \frac{x^{+}_{4,k}-\tilde{x}_{5,j}}{x^{-}_{4,k}-\tilde{x}_{5,j}}\nonumber\\
1&=&\prod^{\frac{L}{2}-1}_{j=1}\frac{\tilde{u}_{5,k}-u_{6,j}+\frac{i}{2}}{\tilde{u}_{5,k}-u_{6,j}-\frac{i}{2}} \prod^{L}_{j=1}\frac{\tilde{x}_{5,k}-x^{+}_{4,j}}{\tilde{x}_{5,k}-x^{-}_{4,j}}\nonumber\\
1&=&\prod^{\frac{L}{2}-1}_{\substack{j=1\\ j \neq k}} \frac{u_{6,k}-u_{6,j}+i}{u_{6,k}-u_{6,j}-i} \prod^{L-2}_{j=1} \frac{u_{6,k}-\tilde{u}_{5,j}-\frac{i}{2}}{u_{6,k}-\tilde{u}_{5,j}+\frac{i}{2}}\, .
\end{eqnarray}
For the highest energy state
we may assume $u_{2,j}=u_{6,j}$ and $u_{3,k}=u_{5,k}$ for $j=1,...,\frac{L}{2}-1$ and $k=1,...,L-2$, respectively. After this is done one notes a striking structural similarity to \eqref{all1}-\eqref{all3}. Apart from the different overall number of magnons the resulting equations differ only by the power of the interaction term in the equation for the $u_4$ roots. It is thus plausible to assume that the $u_5$ and $u_6$ roots will again form stacks, even though this seems to be more difficult to prove in this case. Under this assumption, the derivation of 
the integral equation for the principal density is straightforward and follows the same lines as above. In Fourier space one now gets,
in great similarity to \eqref{finaleq},
\begin{eqnarray}\label{so6equation}
\density(t)&=&\frac{e^t+1}{e^{2\,t}+1}\bigg( \BesselJ_{0}(2gt)\nonumber\\
&&-4\,g^{2}\,t\int^{\infty}_{0}\D t' 
\left ( \frac{e^{t}-1}{e^{t}+1}\,\hat{K}_{0}(2 g t,2 g t')+
\frac{e^{t'}-1}{e^{t'}+1}\,\hat{K}_{1}(2 g t,2 g t')\right)\density(t')
\nonumber\\
&&-4\,g^{2}\,t\int^{\infty}_{0}\D t' 
\left (2\,\hat{K}_n(2 g t,2 g t')+\hat{K}_{d}(2 g t,2 g t')\right)\density(t')\bigg)\,.
\end{eqnarray}
Interestingly, the very same nesting kernel $\hat K_n$
\eqref{Kn} appears, with an overall factor of two, as in the
case of the field-strength operator, {\it cf} \eqref{finaleq}.
This is precisely what one should expect.

It is again rather straightforward to find the weak-coupling 
expansion of the energy to, say, four loops 
(we have highlighted the terms generated by the dressing phase)
\begin{eqnarray}\label{so6eng}
\frac{E(g)}{L}&=&2\,\beta(1)+\zeta_a (1)\nonumber\\
&&-\bigg(4\,\beta(1)\,\beta(2)+6\,\beta(3)+2\,\beta(2)\,\zeta_a(1)-\beta(1)\,\zeta_a(2)-\frac{1}{2}\,\zeta_a(1)\,\zeta_a(2) 
+\frac{3}{4}\,\zeta_a(3) \bigg)g^2\nonumber\\
&&+\bigg 
(8\,\beta(1)\,\beta(2)^2+8\,\beta(2)\,\beta(3)+24\,\beta(1)\,\beta(4)+40\,\beta(5)+4\,\beta(2)^2\,\zeta_a(1)\nonumber\\
&&\;\;\;\;\;+12\,\beta(4)\,\zeta_a(1)-4\,\beta(1)\,\beta(2)\,\zeta_a(2)-2\,\beta(3)\,\zeta_a(2)-2\,\beta(2)\,\zeta_a(1)\,\zeta_a(2)\nonumber\\
&&\;\;\;\;\;+\frac{1}{2}\,\beta(1)\,\zeta_a(2)^2+\frac{1}{4}\,\zeta_a(1)\,\zeta_a(2)^2+\beta(2)\,\zeta_a(3)-\frac{1}{4}\,\zeta_a(2)\,\zeta_a(3)-\frac{3}{2}\,\beta(1)\,\zeta_a(4)\nonumber\\
&&\;\;\;\;\;-\frac{3}{4}\,\zeta_a(1)\,\zeta_a(4)+\frac{5}{4}\,\zeta_a(5) 
\bigg )g^4\nonumber\\
&&-\bigg 
(16\,\beta(1)\,\beta(2)^3+16\,\beta(2)^2\,\beta(3)+16\,\beta(1)\,\beta(3)^2+96\,\beta(1)\,\beta(2)\,\beta(4)+44\,\beta(3)\,\beta(4)\nonumber\\
&&\;\;\;\;\;+40\,\beta(2)\,\beta(5)+200\,\beta(1)\,\beta(6)+350\,\beta(7)+
\bm{32\,\mathbf{\beta}(1)\,\beta(3)\,\zeta(3)}+
8\,\beta(2)^3\,\zeta_a(1)\nonumber\\
&&\;\;\;\;\;+8\,\beta(3)^2\,\zeta_a(1)+48\,\beta(2)\,\beta(4)\,\zeta_a(1)+100\,\beta(6)\,\zeta_a(1)+\bm{16\,\beta(3)\,\zeta_a(1)\,\zeta(3)}\nonumber\\
&&\;\;\;\;\;-12\,\beta(1)\,\beta(2)^2\,\zeta_a(2)-8\,\beta(2)\,\beta(3)\,\zeta_a(2)-24\,\beta(1)\,\beta(4)\,\zeta_a(2)-10\,\beta(5)\,\zeta_a(2)\nonumber\\
&&\;\;\;\;\;-6\,\beta(2)^2\,\zeta_a(1)\,\zeta_a(2)-12\,\beta(4)\,\zeta_a(1)\,\zeta_a(2)+3\,\beta(1)\,\beta(2)\,\zeta_a(2)^2+\beta(3)\,\zeta_a(2)^2\nonumber\\
&&\;\;\;\;\;+\frac{3}{2}\,\beta(2)\,\zeta_a(1)\,\zeta_a(2)^2-\frac{1}{4}\,\beta(1)\,\zeta_a(2)^3-\frac{1}{8}\,\zeta_a(1)\,\zeta_a(2)^3+2\,\beta(2)^2\,\zeta_a(3)\nonumber\\
&&\;\;\;\;\;+4\,\beta(1)\,\beta(3)\,\zeta_a(3)+\frac{11}{2}\,\beta(4)\,\zeta_a(3)+\bm{4\,\beta(1)\,\zeta_a(3)\,\zeta(3)}+2\,\beta(3)\,\zeta_a(1)\,\zeta_a(3)\nonumber\\
&&\;\;\;\;\;+\bm{2\,\zeta_a(1)\,\zeta_a(3)\,\zeta(3)}-\beta(2)\,\zeta_a(2)\,\zeta_a(3)+\frac{1}{8}\,\zeta_a(2)^2\,\zeta_a(3)+\frac{1}{4}\,\beta(1)\,\zeta_a(3)^2\nonumber\\
&&\;\;\;\;\;+\frac{1}{8}\,\zeta_a(1)\,\zeta_a(3)^2-6\,\beta(1)\,\beta(2)\,\zeta_a(4)-\frac{11}{4}\,\beta(3)\,\zeta_a(4)-3\,\beta(2)\,\zeta_a(1)\,\zeta_a(4)\nonumber\\
&&\;\;\;\;\;+\frac{3}{2}\,\beta(1)\,\zeta_a(2)\,\zeta_a(4)+\frac{3}{4}\,\zeta_a(1)\,\zeta_a(2)\,\zeta_a(4)-\frac{11}{32}\,\zeta_a(3)\,\zeta_a(4)+\frac{5}{4}\,\beta(2)\,\zeta_a(5)\nonumber\\
&&\;\;\;\;\;-\frac{5}{16}\,\zeta_a(2)\,\zeta_a(5)-\frac{25}{8}\,\beta(1)\,\zeta_a(6)-\frac{25}{16}\,\zeta_a(1)\,\zeta_a(6)+\frac{175}{64}\,\zeta_a(7) 
\bigg)g^6+\ldots\,.
\end{eqnarray}
For this state, an interesting new set of numbers
appears, namely the Dirichlet $\beta$-function
evaluated at positive integers:
\begin{equation}
\beta(k)=\frac{1}{\Gamma(k)}\int_{0}^{\infty}\frac{dt}{t}\,\frac{t^k\,e^t}{e^{2t}+1}=\sum_{n=0}^{\infty}\frac{(-1)^n}{(2n+1)^k}\,. 
\end{equation}
If $k$ is odd this leads to $\pi^k$ times
rational numbers (related to Euler numbers).
If $k$ is even the numbers $\beta(k)$ cannot be
expressed through $\pi$'s. $\beta(2)$ is Catalan's constant.
All coefficients multiplying the products of
$\zeta$-, $\zeta_a$, and $\beta$-functions  are
integers after factoring out inverse powers of 2.

Note that \eqref{so6equation} is the higher loop generalization of
a one-loop result worked out in \cite{Minahan&Zarembo}.
The latter is of course reproduced from \eqref{so6eng}
since $\beta(1)=\pi/4$ and $\zeta_a(1)=\log(2)$. 
Note that this state's energy satisfies, 
just as the half-filled $\alg{su}(2)$ state's energy 
\eqref{su2eng}, a constant transcendentality principle.

%%%%%%%%%%%%%%%%%%%%%%%%%%%%%%%%%%%%%%%%%%%%%%%%%%%%%%%%%%%%%%%%%%

\section{Outlook}
\label{sec:outlook}

We have shown in detail how to compute the anomalous dimension of 
the field strength pseudo-vacuum \eqref{ops} in the thermodynamic 
limit from the nested asymptotic Bethe equations of \cite{LongRange}. 
Several techniques to reduce the number of equations were introduced 
and a single effective integral equation \eqref{finaleq}
for the distribution density of Bethe roots
was derived from a starting set of five. Combining this equation 
with the expression \eqref{dim} relating the density to the operator
dimension, it is straightforward to find the weak coupling expansion 
of the latter to any desired order, {\it cf} \eqref{dimexp}.
Incidentally, as our equation is analytic in the vicinity of $g=0$, 
it should also be, by analytic continuation, 
just as valid at any value of the coupling. 
It would be very interesting to analyze it in the strong
coupling limit $g \rightarrow \infty$, 
and to interpret the corresponding state in string theory.
The techniques developed in \cite{lecce} might be useful here.

Interestingly, the influence of the nesting on the effective 
Bethe equation
results in a kernel \eqref{Kn} which strongly resembles
the dressing kernel \eqref{Kd} recently proposed in \cite{TransCross}.
This leads us to suggest that the dressing phase should 
originate from the elimination of further, yet to be found
auxiliary Bethe roots. A formal procedure, 
unfortunately plagued with difficulties, 
is briefly discussed in \Appref{app:emulation} 
(see also \cite{Sakai:2007rk}).
The detailed mechanism for how this
happens therefore remains to be worked out. The techniques 
presented in the present paper might prove helpful in this respect.

We would also like to stress that 
``extra Bethe ansatz levels'' in order to improve the asymptotic 
equations of \cite{LongRange} have been proposed
previously. These are natural from general arguments
concerning finite size effects \cite{Ambjorn:2005wa},
from concrete indications that the asymptotic Bethe ansatz needs
to  be corrected \cite{Schafer-Nameki:2006ey}, and
finally from studies indicating that the 
BDS/Hubbard 
magnon dispersion law at strong coupling \cite{Hofman:2006xt}
is to be corrected in a finite volume \cite{Arutyunov:2006gs}
(see also \cite{Astolfi:2007uz} and \cite{Okamura:2006zv}).
In particular, Hubbard-type models are able
to create long-range integrable systems from short
range interactions \cite{hubbard} (see \cite{fioravanti}
for a detailed study on the relation of the nested
to the effective Bethe equations near the antiferromagnetic
vacuum). Furthermore, similar mechanisms have also been
proposed on the level of the string sigma model 
\cite{mann-polchinski,gromov-kazakov}.
It is interesting to note that one key feature of 
\cite{hubbard,mann-polchinski,gromov-kazakov} is that
the BPS vacuum corresponds to a non-trivial
distribution of top-level Bethe roots. This qualitatively 
agrees with our result, which suggests that the 
``physical'' BPS vacuum states \eqref{bps} are created by 
filling up a truly empty ``unphysical'' reference state.
See also the closely related comments in \cite{Klose:2006dd}. 

The number of hidden Bethe roots creating the dressing
factor of \cite{TransCross} is {\it infinite} if the ansatz
is to be asymptotically exact for short operators. This
should be related to the non-compact nature of the
AdS/CFT system. In such a situation Bethe equations
might not necessarily furnish the most effective description,
and an approach based on Sklyanin's separation-of-variables
technique might be more appropriate. This would then
replace the Bethe equations by functional equations
for a nested set of Baxter-Q functions. Some of the latter
should be {\it non-polynomial} in nature. 
Possibly the techniques used in 
\cite{bytsko-teschner,Belitsky:2006cp} might be useful in this context.
 
\subsection*{Acknowledgments}
We would like to thank Niklas Beisert, Burkhard Eden,
Lisa Freyhult, Sergey Frolov, Didina Serban, J\"org Teschner 
and Arkady Tseytlin for useful discussions.

%%%%%%%%%%%%%%%%%%%%%%%%%%%%%%%%%%%%%%%%%%%%%%%%%%%%%%%%%%%%%%%%%%
\appendix

\section{Properties of the Generating Polynomials}\label{app:polyprop}
In this section we will analyze some properties of the generating polynomials $Q_4$ and $Q_6$. The mathematical description of the formation of stacks will be given in the next section.

Let us start with the following observation: if $Q(x)$ is a polynomial with only real roots then
\begin{displaymath}
S(x)=Q(x+i\ s)+\alpha \ Q(x-i\ s)
\end{displaymath}
with $|\alpha|=1$ also has only real roots. To prove this one observes, that for any real $y$
\begin{displaymath}
|Q(x+i y)|=|Q(x-iy)|=\prod^{n}_{i=1} |(x-w_j)+iy|
\end{displaymath}
 grows monotonically with increasing $|y|$. Furthermore, if $x_j$ satisfies:
\begin{displaymath}
S(x_j)=Q(x_j+i\ s)+\alpha \ Q(x_j-i\ s)=0
\end{displaymath}
then $|Q(x_j+i\ s)|=|Q(x_j -i\ s)|$. Together with the previous 
observation this implies $x_j \in \mathbb{R}$. 
Applying this theorem twice to $w^L$ one proves immediately 
that $Q_6(w)$ has only real roots. Similarly
\begin{displaymath}
Q_4(u)=W(u+\frac{i}{2})-W(u-\frac{i}{2}), \qquad W(u)=u^L ((u+i)^L-(u-i)^L)
\end{displaymath}
also has only real roots.
If one orders the roots of $Q(x)$ as follows $x_1 \leq x_2 \leq ... \leq x_n$ one finds that
\begin{displaymath}
Q'(x_i)\ Q'(x_{i+1}) \leq 0\,.
\end{displaymath}
Thus, either $x_i$ or $x_{i+1}$ are zeros of $Q'(x)$ or there 
exist a $\theta$ with $0<\theta<1$ such that 
\begin{displaymath}
Q'(x_i+\theta (x_{i+1}-x_{i}))=0, \qquad i=1,...,n-1\,.
\end{displaymath}
In case of $Q_6(w)$ one finds
\begin{displaymath}
\frac{\D}{\D w}Q_6 (w,L)=L\ Q_6 (w,L-1)\,.
\end{displaymath}
Thus, between any two zeros of $Q_6(w,L)$ there lies a zero of $Q_6(w,L-1)$. 
Furthermore it is fairly easy to prove that the resultant (given by the 
determinant of the corresponding Sylvester matrix) of $Q_6(w,L)$ 
and $Q_6(w,L-1)$ is always non-vanishing.
Finally, it is possible to find an approximate formula for the extreme 
roots of $Q_6 (w,L)$. Expecting this roots to scale with $L$ we set
\begin{displaymath}
w_{max}=\pm aL\,.
\end{displaymath}
Using
\begin{displaymath}
\lim_{L \to \infty} (1+\frac{c}{L})^L=e^{c}
\end{displaymath}
we write
\begin{displaymath}
Q_6(\pm a L,L)=-8 (\pm a L)^L \cos \bigg (\frac{\pm 1}{2 a}\bigg) \sin \bigg(\frac{\pm 1}{2 a}\bigg)^{2}\,+\mathcal{O}(\frac{1}{L}).
\end{displaymath}
Setting $a=\frac{1}{\pi}$ one finds that $Q_6(\frac{1}{\pi}L +\epsilon ,L)$ 
changes sign with $\epsilon$ and hence,
\begin{equation}\label{exroot}
w_{max}\simeq \pm \frac{1}{\pi} L\,, \qquad L>>1\,.
\end{equation}
We conclude that for any $L > 2$ the generating polynomial $Q_6(w,L)$  
has $(L-2)$ distinct real roots, forming a dense set in $\mathbb{R}$ for $L \to \infty$.
Curiously $Q_6 (w,L)$ is the polynomial solution of
\begin{displaymath}
Q_6 (w+\frac{3 i}{2})+Q_6 (w-\frac{3 i}{2})+Q_6(w+\frac{i}{2})+Q_6(w-\frac{i}{2})= \ 2^L \ Q_6 (\frac{w}{2})
\end{displaymath}
This should be interpreted as the Baxter equation for the $u_6$ roots.
Similarly one can find extreme roots of $Q_4\ (u,L)$ and ${Q}_5 \ (v,L)$. 
Surprisingly they are also given by (\ref{exroot}).
%
%%%%%%%%%%%%%%%%%%%%%%%%%%%%%%%%%%%%%%%%%%%%%%%%%%%%%%%%%%%%%%%%%%%%%%%%%%%%%%%%%%%%%%%%
%
\section{Formation of Stacks}\label{app:stacks}
Extreme roots of $Q_5$ are indeed real (as demonstrated above). In this 
section we will, however, prove that almost all roots of $Q_5(v,L)$, 
for large values of L, occupy two contours shifted from the real axis 
by $\pm \frac{i}{2}$, respectively. In other words we will show that
\begin{equation}
Q_{5}(x,L)=Q_{6}(x+\frac{i}{2},L)\,Q_{6}(x-\frac{i}{2},L)+\mathcal{O}(\frac{1}{L})
\end{equation}
where $\mathcal{O}(\frac{1}{L})$ can be neglected in the large $L$ limit.
%%%%%%%%%%%%%%%%%%%%%%%%%%%%%%%%%%%%%%%%%%%%%%%%%%%
%
\subsection{Power Expansion of the Polynomials}
After some computation it is possible to find for the polynomial $Q_5(v)$
\begin{eqnarray}
Q_5 (v)&=&3 v^{2L}-2 \sum^{L}_{k=0} \binom{L}{k} v^{2L-k} \cos \frac{\pi k}{2}-4\sum^{L}_{k=0} 2^{k} \binom{L}{k}v^{2L-k} \cos \frac{\pi k}{2}\nonumber\\
&&+2\sum^{L}_{k=0}\binom{L}{k} v^{2L-k}\cos\frac{\pi k}{2} \ {_2 F _1} [-L,-k;1+L-k;2]\nonumber\\
&&+2\sum^{2L}_{k=L+1}\binom{L}{2L-k}2^{k-L} v^{2L-k}\cos\frac{\pi k}{2} \ {_2 F _1} [-L,k-2L;1+k-L;2]\nonumber\\
&&+\sum^{L}_{k=0}\binom{L}{k} v^{2(L-k)} 4^k.
\end{eqnarray}
Thus one finds immediately that 
\begin{displaymath}
Q_5 (v)=\sum^{2L-4}_{n=0} c_n v^n
\end{displaymath}
where the coefficients are given by
\begin{eqnarray}
c_n=&-&2\ \binom{L}{2L-n}\cos \frac{\pi (2L-n)}{2} -4 \ 2^{2L-n} \binom{L}{2L-n} \cos \frac{\pi (2L-n)}{2}\nonumber\\
&+&2\binom{L}{2L-n} \cos \frac{\pi (2L-n)}{2} {_2 F_1}[-L,n-2L;1+n-L;2]+\binom{L}{n/2} \cos^2 \left(\frac{\pi n}{2}\right)4^{L-\frac{n}{2}}\nonumber\\
\end{eqnarray}
for $n=L,L+1,...,2L-4$ and 
\begin{eqnarray}
c_n=&-&2\binom{L}{2L-n} \cos \frac{\pi (2L-n)}{2} -4 \ 2^{2L-n} \binom{L}{2L-n} \cos \frac{\pi (2L-n)}{2}\nonumber\\
&+&2\binom{L}{n} 2^{L-n}\cos \frac{\pi (2L-n)}{2} {_2 F_1}[-L,-n;1+L-n;2]+\binom{L}{n/2} \cos^2\left( \frac{\pi n}{2}\right)4^{L-\frac{n}{2}}\nonumber\\
\end{eqnarray}
for $n=0,...,L-1$. Similarly one finds
\begin{eqnarray}
Q_6 (w+\frac{i}{2}) Q_6 (w-\frac{i}{2})&=&w^{2L}-2\sum^{2L}_{k=0} \binom{2L}{k} w^{2L-k} \cos \frac{\pi k}{2}-2\sum^{L}_{k=0}2^k \binom{L}{k} w^{2L-k} \cos \frac{\pi k}{2}\nonumber\\
&+&2\sum^{L}_{k=0} \binom{L}{k} w^{2(L-k)}+\sum^{L}_{k=0} \binom{L}{k} w^{2(L-k)}4^k\nonumber\\
&+&2\sum^{L}_{k=0} w^{2L-k} \binom{L}{k} \cos \frac{\pi k}{2}\bigg( {_2 F_1}[-L,-k;1+L-k;2]\nonumber\\
&-&{_2 F_1}[-L,-k;1+L-k;-2] \bigg)\nonumber\\
&+&2\sum^{2L}_{k=L+1} w^{2L-k} 2^{k-L}\binom{L}{2L-k} \cos \frac{\pi k}{2}\bigg( {_2 F_1}[-L,k-2L;1+k-L;2]\nonumber\\
&-&{_2 F_1}[-L,k-2L;1+k-L;-2] \bigg)\,.
\end{eqnarray}
The corresponding coefficients in the power expansion of
\begin{displaymath}
Q_6 (w+\frac{i}{2}) Q_6 (w-\frac{i}{2})=\sum^{2L-4}_{n=0} d_n w^n
\end{displaymath}
can easily be read off 
\begin{eqnarray}
d_n&=&-2\binom{2L}{n} \cos \frac{\pi (2L-n)}{2}-2\ 2^{2L-n} \binom{L}{2L-n} \cos \frac{\pi (2L-n)}{2}\nonumber\\
&&+2 \binom{L}{n/2} \cos^2\left( \frac{\pi n}{2}\right)+\binom{L}{n/2} 4^{L-\frac{n}{2}} \cos^2 \left( \frac{\pi n}{2}\right)\nonumber\\
&&+2 \binom{L}{2L-n} \cos \frac{\pi (2L-n)}{2} \bigg( {_2 F _1}[-L,n-2L;1+n-L;2]\nonumber\\
&&-{_2 F _1}[-L,n-2L;1+n-L;-2] \bigg)
\end{eqnarray}
for $n=L,...,2L-4$ and
\begin{eqnarray}
d_n&=&-2\binom{2L}{n} \cos \frac{\pi (2L-n)}{2}-2\ 2^{2L-n} \binom{L}{2L-n} \cos \frac{\pi (2L-n)}{2}\nonumber\\
&&+2 \binom{L}{n/2}\cos^2 \left(\frac{\pi n}{2}\right)+
\binom{L}{n/2} 4^{L-\frac{n}{2}} \cos^2 \left(\frac{\pi n}{2}\right)\nonumber\\
&&+2\ 2^{L-n} \binom{L}{n} \cos \frac{\pi (2L-n)}{2} \bigg( {_2 F _1}[-L,-n;1+L-n;2]\nonumber\\
&&-(-1)^{L-n} {_2 F _1}[-L,-n;1+L-n;-2] \bigg)
\end{eqnarray}
for $n=0,1,...,L-1$\,. 
%
%%%%%%%%%%%%%%%%%%%%%%%%%%%%%%%%%%%%%%%%%%%%%%%%
\subsection{Emergence of the Stack Picture}
In this subsection we assume $n \geq L$ ($n<L$ can be analyzed analogously). 
Furthermore without loss of generality we consider $L$ and $n$ to be even. 
The non-zero coefficients are then given by
\begin{equation}\label{c}
c_n(L) = 2 (-1)^{\frac{n}{2}} \binom{L}{2L-n}\bigg({_2 F_1}[-L,n-2L;1+n-L;2]-1-2\ 2^{2L-n}\bigg)+\binom{L}{n/2} 4^{L-\frac{n}{2}}
\end{equation}
\begin{eqnarray}\label{d}
d_n(L)&=&2 (-1)^{\frac{n}{2}} \binom{L}{2L-n} \bigg({_2 F_1}[-L,n-2L;1+n-L;2]\nonumber\\
&&-{_2 F_1}[-L,n-2L;1+n-L;-2]-2^{2L-n}\bigg)\nonumber\\
&&-2(-1)^{\frac{n}{2}}\ \binom{2L}{n}+\binom{L}{n/2}\bigg(2+ \ 4^{L-\frac{n}{2}}\bigg)\,.
\end{eqnarray}
%
%%%%%%%%%%%%%%%%%%%%%%%%%%%%%%%%%%%%%%%%%%%%%%
\subsubsection{\texorpdfstring{${\mathbf {n=2L-a}}$}{} case}
The case $n=2L-a$ with $a\geq 4$ and finite $a$ is fairly easy 
to analyze. Noting that
\begin{displaymath}
\frac{(-L)_k}{(1+L-a)_k}=(-1)^k \bigg(1+\frac{ak(1-k)}{L}+\frac{1}{2L^2} (a-k)k(-1+a+ak-k^2)+...\bigg)
\end{displaymath}
one can immediately shown that for the large $L$ expansion of ${_2 F_1}[-L,-a;1+L-a;\pm 2]$
\begin{eqnarray}
{_2 F_1}[-L,-a;1+L-a;2]&=&3^a \bigg(1+\frac{2}{9L} \ a(a-1)+\frac{2}{81 L^2}\ a(a^3+3a^2-7a+3)+...\bigg)\nonumber\\
{_2 F_1}[-L,-a;1+L-a;-2]&=&(-1)^a \bigg(1-\frac{2}{L} \ a(a-1)+\frac{2}{L^2}a (a^3-7a^2+13a-7)+...\bigg)\,.\nonumber
\end{eqnarray}
Expanding $c_n$ and $d_n$ we note that
\begin{displaymath}
\frac{c_n(L)}{d_n(L)}=1+{\mathcal O}(\frac{1}{L})
\end{displaymath}
for large values of $L$.
%
%%%%%%%%%%%%%%%%%%%%%%%%%%%%%%%%%%%%%%%%%%%%%%%
%
\subsubsection{Other values of \texorpdfstring{$\mathbf{n}$}{}}
Other values of $n$ are much more difficult to analyze since their generic dependence on $L$ is
\begin{displaymath}
n=\alpha L+\beta\,, \qquad 1\leq \alpha <2
\end{displaymath}
and the summation limit of ${_2 F_1}$ depends now on $L$. Here we will 
show that $\frac{c_n}{d_n}-1$ is at least of order $\mathcal {O}(\frac{1}{L})$.
We start with noting that for this values of $n$ one has
\begin{displaymath}
{_2 F _1} [-L,n-2 L;1+n-L;2]>1+2^{2L-n}+\frac{L (n-2L)}{L-n-1}(2+ 2^{2L-n-1})+...
\end{displaymath}
i.e.~terms of the form $\binom{L}{2L-n}\ 2^{2L-n}$ are least of all sub-leading.
Next we observe that
\begin{displaymath}
\binom{L}{2L-n}\  {_2 F _1} [-L,n-2 L;1+n-L;2]= \sum^{L}_{j=0} \binom{L}{j} \binom{L}{2L-n-j} 2^j \ >\end{displaymath}
\begin{displaymath}
>\sum^{L}_{j=0} \binom{L}{j} \binom{L}{2L-n-j}\  j \ = L \frac{n}{2L} \binom{2L}{n}\  \simeq \ \alpha L\  \binom{2L}{n}+\mathcal{O}(\frac{1}{L})\,.
\end{displaymath}
This implies that the $\binom{2L}{n}$ term in (\ref{d}) is at least sub-leading.
Let us define the function
\begin{displaymath}
f_{2L-n,L}(x)=\sum^{L}_{j=0} \binom{L}{j} \binom{L}{2L-n-j} x^j=P^{(L-\tilde{n},-2L-1)}_{\tilde{n}} \bigg(1-2x\bigg)
\end{displaymath}
where $P^{(\alpha,\beta)}_{m}(x)$ stands for the Jacobi polynomials 
(see e.g.~\cite{Szego}) and $\tilde{n}=2L-n$.
It is easy to prove the following properties of $f_{n,L}(x)$
\begin{equation}\label{rel}
x^n f_{n,L}\bigg(\frac{1}{x}\bigg)=f_{n,L}(x)\,, \qquad x^{n-L} f_{2L-n,L}(x)=f_{n,L}(x)\,.
\end{equation}
A straightforward application of these is the relation
\begin{displaymath}
f_{2L-n,L}(x)=x^L f_{n,L}\bigg(\frac{1}{x}\bigg)
\end{displaymath}
which can be used to get an upper bound
\begin{displaymath}
\binom{L}{2L-n} {_2 F_1}[-L,n-2L;1+n-L;2]\ <\ 2^L \ \binom{2L}{n}.
\end{displaymath}
$f_{2L-n,L}(x)$ is a polynomial of degree $2L-n$ and thus has $2L-n$ roots. 
Because of Descartes rule and the properties of Jacobi polynomials all the 
zeros are real, distinct and negative. Furthermore because of (\ref{rel})
\begin{displaymath}
f_{2L-n,L}(x)=c\prod^{L-\frac{n}{2}}_{j=1} \bigg(x+|x_j|\bigg)\bigg(x+\frac{1}{|x_j|}\bigg)
\end{displaymath}
where all $x_j \in (-1,0)$. We thus can write\footnote{For simplicity assume $x_j \neq -\frac{1}{2}$.}
\begin{displaymath}
\left| \frac{f_{2L-n,L}(2)}{f_{2L-n,L}(-2)} \right|=\left| \frac{{_2 F _1}[-L,n-2L;1+n-L;2]}{{_2 F _1}[-L,n-2L,1+n-L,-2]} \right|=\prod^{L-\frac{n}{2}}_{j=1} \left|\frac{|x_j|+2}{|x_j|-2}\right|\ \left| \frac{\frac{1}{|x_j|}+2}{\frac{1}{|x_j|}-2} \right|.
\end{displaymath}
If the roots are densely distributed the right hand side grows exponentially with growing $L$. To say something more about the root distribution we observe that
\begin{displaymath}
f_{2L-n,L} (x)= P^{(L-\tilde{n},-2L-1)}_{\tilde{n}} (1-2x)=(1-x)^{2L-n}\  P^{n-L,n-L}_{2L-n} \bigg(\frac{x+1}{1-x}\bigg).
\end{displaymath}
Since $P^{(\alpha,\alpha)}_{m}(y)$ with $\alpha>-1$ has roots densely distributed on $(-1,1)$ we conclude that the roots of $f_{2L-n,L}(x)$ are densely distributed everywhere on $(-\infty,0)$ which completes the proof.
%
%%%%%%%%%%%%%%%%%%%%%%%%%%%%%%%%%%%%%%%%%%%%%%%%%%%%%%%%%%%%%%%%%%%%%%%%%%%%%%%%%%%%%%%%
%
\section{Exact Results for Finite \texorpdfstring{$L$}{}}
\label{app:numres}
Here we present the energy  for the field strength operators \eqref{ops} of 
finite length $L$ in powers of the coupling constant. 
Their anomalous dimension is \eqref{dimeng}.
The results were obtained using {\tt{Mathematica}}. 
The $\zeta$-functions are exclusively generated by the dressing factor.\\
\textbf{L=3\footnote{Please note that for this particular ``short''
state wrapping effects compete with the deformation caused by the dressing 
factor. The Bethe equations are not reliable anymore, and we
thus omit the $\Op(g^6)$ term.}}:
\begin{displaymath}
E_3= 9-\frac{81}{2}g^2+315g^4+\ldots
\end{displaymath}
\textbf{L=4}:
\begin{displaymath}
E_4=12-50g^2+\frac{1515}{4}g^4-\bigg(\frac{513937}{144} +216\, \zeta(3)\bigg)g^6
\end{displaymath}
\begin{displaymath}
+\bigg(\frac{22129823}{576}+3302\, \zeta(3)+2160\, \zeta(5) \bigg)g^8+\ldots
\end{displaymath}
\textbf{L=5}:
\begin{displaymath}
E_5=15-\frac{129}{2}g^2+\frac{39411}{80}g^4-\bigg(\frac{7346253}{1600}+270\, \zeta(3)\bigg)g^6
\end{displaymath}
\begin{displaymath}
+\bigg(\frac{1539949881}{32000}+\frac{8307\,\zeta(3)}{2}+2700\, \zeta(5)\bigg)g^8+\ldots
\end{displaymath}
\textbf{L=6}:
\begin{displaymath}
E_6=18-\frac{837}{11}g^2+\frac{6278355}{10648}g^4-\bigg(\frac{29266837713}{5153632}+324\, \zeta(3) \bigg)g^6
\end{displaymath}
\begin{displaymath}
+\bigg(\frac{76857234976107}{1247178944}+\frac{605205\, \zeta(3)}{121}+3240\,\zeta(5)\bigg)g^8+\ldots
\end{displaymath}
\textbf{L=7}:
\begin{displaymath}
E_7=21-\frac{179}{2}g^2+\frac{76603}{112}g^4-\bigg(\frac{181131695}{28224}+378\, \zeta(3)\bigg)g^6
\end{displaymath}
\begin{displaymath}
+\bigg(\frac{8959397257}{131712}+\frac{11641\,\zeta(3)}{2}+3780\,\zeta(5)\bigg)g^8+\ldots
\end{displaymath}
\textbf{L=8}:
\begin{displaymath}
E_{8}=24-\frac{4380}{43}g^2+\frac{125533809}{159014}g^4-\bigg(\frac{8840715968859}{1176067544}+432\,\zeta(3)\bigg)g^6
\end{displaymath}
\begin{displaymath}
+\bigg(\frac{346753221469919673}{4349097777712}+\frac{12323484\,\zeta(3)}{1849}+4320\,\zeta(5)\bigg)g^8+\ldots
\end{displaymath}

There are several conclusions one can draw from these exact
results as compared to \eqref{dimexp}. It is clear that the higher-loop 
corrections are not exactly proportional to $L$ for finite length operators. 
This is presumably caused by length changing processes, see \cite{LongRange}. 
Also, the transcendentality principle is violated. The terms of highest 
transcendentality, however,  
are strictly proportional to $L$ and precisely agree with the ones
derived  in \eqref{dimexp}. 

%%%%%%%%%%%%%%%%%%%%%%%%%%%%%%%%%%%%%%%%%%%%%%%%%%%%%%%%%%%%%%%%%%
\section{Emulation of the Dressing Phase?}\label{app:emulation}

In this paper we have argued that a nested Bethe ansatz naturally 
gives rise to a convolution structure very similar to the one in
the proposed all-loop dressing phase.
In this appendix\footnote{
The results of \Appref{app:emulation} were known
to us for some time, but we did not include them in an
earlier version of this paper due to the various difficulties 
discussed below.
While preparing an improved version of this article similar ideas appeared in \cite{Sakai:2007rk}. We therefore decided
to include our notes in order to discuss the problems
which would have to be resolved in order to make this 
and similar mechanisms viable.
Furthermore, our procedure differs in several 
ways from \cite{Sakai:2007rk}. In particular, we 
distinguish nesting from dressing roots (there is no
direct interaction between them), and we do not
assume a macroscopic number of momentum carrying main
roots.}
we will ask whether we can modify or supplement the asymptotic Bethe equations of \cite{LongRange} in order to
construct a well-defined set of equations leading to the
correct dressing phase.

We notice that the difference between the nesting phase
\eqref{Kn} and the dressing phase \eqref{Kd} is just,
apart from a factor of two, a minus sign in the diagonal kernel
$1/(e^t \pm 1)$ inside the convolution. The factor of two is
easily fixed by using two instead of one set of auxiliary
roots, just as in the case of the $\alg{so}(6)$ nesting,
see \eqref{so6equation}. We claim that the sign may also be 
{\it formally} fixed by emulating the full set of asymptotic
Bethe equations {\it with} dressing factor by the 
following set of equations:
\begin{eqnarray}\label{MAIN}
1&=&
\left ( \frac{x^{+}_{4,k}}{x^{-}_{4,k}} \right )^{2M_d-L}\,
\prod_{\substack{j=1 \\ j \neq k}}^{K_4}
\left( \frac{x^{+}_{4,k}-x^{-}_{4,j}}{x^{-}_{4,k}-x^{+}_{4,j}}\,
\frac{1-g^2/x^{+}_{4,k}x^{-}_{4,j}}{1-g^2/x^{-}_{4,k}x^{+}_{4,j}}\right )\nonumber\\
&&\times
\prod_{j=1}^{K_1}
\frac{1-g^{2}/x_{4,k}^{-}x^{}_{1,j}}{1-g^{2}/x_{4,k}^{+}x^{}_{1,j}}
\prod_{j=1}^{K_3}
\frac{x_{4,k}^{-}-x^{}_{3,j}}{x_{4,k}^{+}-x^{}_{3,j}}
\prod_{j=1}^{K_5}
\frac{x_{4,k}^{-}-x^{}_{5,j}}{x_{4,k}^{+}-x^{}_{5,j}}
\prod_{j=1}^{K_7}
\frac{1-g^{2}/x_{4,k}^{-}x^{}_{7,j}}{1-g^{2}/x_{4,k}^{+}x^{}_{7,j}}\nonumber\\
&&\times
\prod_{j=1}^{M_{d}}
\frac{1-g^{2}/x_{4,k}^{-}x^{}_{\bar{1},j}}{1-g^{2}/x_{4,k}^{+}x^{}_{\bar{1},j}}
\prod_{j=1}^{M_{d}}
\frac{x_{4,k}^{-}-x^{}_{\bar{3},j}}{x_{4,k}^{+}-x^{}_{\bar{3},j}}
\prod_{j=1}^{M_{d}}
\frac{x_{4,k}^{-}-x^{}_{\bar{5},j}}{x_{4,k}^{+}-x^{}_{\bar{5},j}}
\prod_{j=1}^{M_{d}}
\frac{1-g^{2}/x_{4,k}^{-}x^{}_{\bar{7},j}}{1-g^{2}/x_{4,k}^{+}x^{}_{\bar{7},j}}\, ,\nonumber\\
\end{eqnarray}
\begin{eqnarray}\label{FULL SET1}
1&=&\prod_{j=1}^{M_{d}}
\frac{u_{\bar{1},k}-u_{\bar{2},j}+\frac{i}{2}}{u_{\bar{1},k}-u_{\bar{2},j}-\frac{i}{2}}
\prod_{j=1}^{K_4}
\frac{1-g^{2}/x_{\bar{1},k}x_{4,j}^{+}}{1-g^{2}/x_{\bar{1},k}x_{4,j}^{-}}\, ,\nonumber\\
1&=&\prod_{\substack{j=1\\j \neq k}}^{M_{d}}
\frac{u_{\bar{2},k}-u_{\bar{2},j}-i}{u_{\bar{2},k}-u_{\bar{2},j}+i}
\prod_{j=1}^{M_{d}}
\frac{u_{\bar{2},k}-u_{\bar{3},j}+\frac{i}{2}}{u_{\bar{2},k}-u_{\bar{3},j}-\frac{i}{2}}
\prod_{j=1}^{M_{d}}
\frac{u_{\bar{2},k}-u_{\bar{1},j}+\frac{i}{2}}{u_{\bar{2},k}-u_{\bar{1},j}-\frac{i}{2}}\, ,\nonumber\\
1&=&\prod_{j=1}^{M_{d}}
\frac{u_{\bar{3},k}-u_{\bar{2},j}+\frac{i}{2}}{u_{\bar{3},k}-u_{\bar{2},j}-\frac{i}{2}}
\prod_{j=1}^{K_4}
\frac{x_{\bar{3},k}-x_{4,j}^{+}}{x_{\bar{3},k}-x_{4,j}^{-}}\, ,\nonumber\\
\end{eqnarray}
\begin{eqnarray}\label{FULL SET2}
1&=&\prod_{j=1}^{K_2}
\frac{u_{1,k}-u_{2,j}+\frac{i}{2}}{u_{1,k}-u_{2,j}-\frac{i}{2}}
\prod_{j=1}^{K_4}
\frac{1-g^{2}/x^{}_{1,k}x_{4,j}^{+}}{1-g^{2}/x^{}_{1,k}x_{4,j}^{-}}\, ,\nonumber\\
1&=&\prod_{\substack{j=1\\j \neq k}}^{K_2}
\frac{u_{2,k}-u_{2,j}-i}{u_{2,k}-u_{2,j}+i}
\prod_{j=1}^{K_3}
\frac{u_{2,k}-u_{3,j}+\frac{i}{2}}{u_{2,k}-u_{3,j}-\frac{i}{2}}
\prod_{j=1}^{K_1}
\frac{u_{2,k}-u_{1,j}+\frac{i}{2}}{u_{2,k}-u_{1,j}-\frac{i}{2}}\, ,\nonumber\\
1&=&\prod_{j=1}^{K_2}
\frac{u_{3,k}-u_{2,j}+\frac{i}{2}}{u_{3,k}-u_{2,j}-\frac{i}{2}}
\prod_{j=1}^{K_4}
\frac{x^{}_{3,k}-x_{4,j}^{+}}{x^{}_{3,k}-x_{4,j}^{-}}\, ,\nonumber\\
%%
%MAIN
%%
\end{eqnarray}
\begin{eqnarray}\label{FULL SET3}
1&=&\prod_{j=1}^{K_6}
\frac{u_{5,k}-u_{6,j}+\frac{i}{2}}{u_{5,k}-u_{6,j}-\frac{i}{2}}
\prod_{j=1}^{K_4}
\frac{x^{}_{5,k}-x_{4,j}^{+}}{x^{}_{5,k}-x_{4,j}^{-}}\, ,\nonumber\\
1&=&\prod_{\substack{j=1\\j \neq k}}^{K_6}
\frac{u_{6,k}-u_{6,j}-i}{u_{6,k}-u_{6,j}+i}
\prod_{j=1}^{K_5}
\frac{u_{6,k}-u_{5,j}+\frac{i}{2}}{u_{6,k}-u_{5,j}-\frac{i}{2}}
\prod_{j=1}^{K_7}
\frac{u_{6,k}-u_{7,j}+\frac{i}{2}}{u_{6,k}-u_{7,j}-\frac{i}{2}}\, ,\nonumber\\
1&=&\prod_{j=1}^{K_6}
\frac{u_{7,k}-u_{6,j}+\frac{i}{2}}{u_{7,k}-u_{6,j}-\frac{i}{2}}
\prod_{j=1}^{K_4}
\frac{1-g^{2}/x^{}_{7,k}x_{4,j}^{+}}{1-g^{2}/x^{}_{7,k}x_{4,j}^{-}}\, ,\nonumber\\
\end{eqnarray}
\begin{eqnarray}\label{FULL SET4}
1&=&\prod_{j=1}^{M_{d}}
\frac{u_{\bar{5},k}-u_{\bar{6},j}+\frac{i}{2}}{u_{\bar{5},k}-u_{\bar{6},j}-\frac{i}{2}}
\prod_{j=1}^{K_4}
\frac{x_{\bar{5},k}-x_{4,j}^{+}}{x_{\bar{5},k}-x_{4,j}^{-}}\, ,\nonumber\\
1&=&\prod_{\substack{j=1\\j \neq k}}^{M_{d}}
\frac{u_{\bar{6},k}-u_{\bar{6},j}-i}{u_{\bar{6},k}-u_{\bar{6},j}+i}
\prod_{j=1}^{M_{d}}
\frac{u_{\bar{6},k}-u_{\bar{5},j}+\frac{i}{2}}{u_{\bar{6},k}-u_{\bar{5},j}-\frac{i}{2}}
\prod_{j=1}^{M_{d}}
\frac{u_{\bar{6},k}-u_{\bar{7},j}+\frac{i}{2}}{u_{\bar{6},k}-u_{\bar{7},j}-\frac{i}{2}}\, ,\nonumber\\
1&=&\prod_{j=1}^{M_{d}}
\frac{u_{\bar{7},k}-u_{\bar{6},j}+\frac{i}{2}}{u_{\bar{7},k}-u_{\bar{6},j}-\frac{i}{2}}
\prod_{j=1}^{K_4}
\frac{1-g^{2}/x_{\bar{7},k}x_{4,j}^{+}}{1-g^{2}/x_{\bar{7},k}x_{4,j}^{-}}\, .
\end{eqnarray}
These equations are very similar to the ones in Table 5 of
\cite{LongRange}. However, the dressing factor 
$\prod\,\sigma^2$ of Table 5 is now missing and is
supposed to be emulated by
\begin{equation}
\sigma^2_k(x_{4,1}^{},\ldots,x_{4,K_4}^{})=
\prod_{j=1}^{M_{d}}
\frac{x_{4,k}^{-}-g^{2}/x^{}_{\bar{1},j}}{x_{4,k}^{+}-g^{2}/x^{}_{\bar{1},j}}
\,
\frac{x_{4,k}^{-}-x^{}_{\bar{3},j}}{x_{4,k}^{+}-x^{}_{\bar{3},j}}\,
\frac{x_{4,k}^{-}-x^{}_{\bar{5},j}}{x_{4,k}^{+}-x^{}_{\bar{5},j}}\,
\frac{x_{4,k}^{-}-g^{2}/x^{}_{\bar{7},j}}{x_{4,k}^{+}-g^{2}/x^{}_{\bar{7},j}}
\, .
\end{equation}
Note that we have introduced an additional set of 
$2 \times 3 \times M_d$
auxiliary ``dressing" roots denoted by a bar on the flavor indices
$\{\bar{1},\bar{2},\bar{3}\},\{\bar{5},\bar{6},\bar{7}\}$.
They should not be confused with the set of $3+1+3$ ``nesting" 
Bethe roots, carrying unbarred indices, with multiplicities 
$(K_1,K_2,K_3,K_4,K_5,K_6,K_7)$. These are identical to the 
ones in \cite{LongRange}. We will now prove our claim,
which however requires the $M_d \rightarrow \infty$ limit.
We will also show, as required by consistency, that the emulated
dressing  phase indeed {\it factorizes}
\begin{equation}\label{factor}
\sigma^2_k(x_{4,1}^{},\ldots,x_{4,K_4}^{})=
\prod_{\substack{j=1\\ j \neq k}}^{K_4}
\sigma^2(x_{4,k}^{},x_{4,j}^{})\, ,
\end{equation}
as required by the asymptotic Bethe equations of Table 5
in \cite{LongRange}. Notice that there are various 
ways to rewrite the equations  \eqref{MAIN} - \eqref{FULL SET4}
since we may always perform dynamic transformations 
\cite{LongRange} and therewith 
trade back and forth, respectively, roots of type 
$\bar{1},\bar{7}$ and type $\bar{3},\bar{5}$. 

The derivation of the dressing factor formally proceeds as in the
main body of this article. We will assume the same mechanism
of stack formation. We may then eliminate the dressing roots of
type $\{\bar{1},\bar{3}\}$ and $\{\bar{5},\bar{7}\}$ and
write ``effective'' equations coupling the dressing roots
of type $\bar{2}$ and $\bar{6}$ to the momentum carrying
roots of type $4$. This yields 
\begin{eqnarray}\label{dressing1}
1&=&\prod_{\substack{j=1\\ j \neq k}}^{M_{d}}
\frac{u_{\bar{2},k}-u_{\bar{2},j}-i}{u_{\bar{2},k}-u_{\bar{2},j}+i}
\prod_{j=1}^{K_4}\frac{1-g^{2}/x_{\bar{2},k}^{+}x_{4,j}^{-}}{1-g^{2}/x_{\bar{2},k}^{-}x_{4,j}^{+}}\,
\frac{1-g^{2}/x_{\bar{2},k}^{-}x_{4,j}^{-}}{1-g^{2}/x_{\bar{2},k}^{+}x_{4,j}^{+}}\, ,
\\
\label{dressing main}
\sigma^2_k(x_{4,1}^{},\ldots,x_{4,K_4}^{})&=&
\prod_{j=1}^{M_{d}}\frac{1-g^{2}/x_{4,k}^{-}x_{\bar{2},j}^{+}}{1-g^{2}/x_{4,k}^{+}x_{\bar{2},j}^{-}}\,
\frac{1-g^{2}/x_{4,k}^{-}x_{\bar{2},j}^{-}}{1-g^{2}/x_{4,k}^{+}x_{\bar{2},j}^{+}}\nonumber\\
&&\times
\prod_{j=1}^{M_{d}}\frac{1-g^{2}/x_{4,k}^{-}x_{\bar{6},j}^{+}}{1-g^{2}/x_{4,k}^{+}x_{\bar{6},j}^{-}}\,
\frac{1-g^{2}/x_{4,k}^{-}x_{\bar{6},j}^{-}}{1-g^{2}/x_{4,k}^{+}x_{\bar{6},j}^{+}}\, ,
\\
\label{dressing 2}
1&=&\prod_{\substack{j=1\\ j \neq k}}^{M_{d}}
\frac{u_{\bar{6},k}-u_{\bar{6},j}-i}{u_{\bar{6},k}-u_{\bar{6},j}+i}
\prod_{j=1}^{K_4}\frac{1-g^{2}/x_{\bar{6},k}^{+}x_{4,j}^{-}}{1-g^{2}/x_{\bar{6},k}^{-}x_{4,j}^{+}}\,
\frac{1-g^{2}/x_{\bar{6},k}^{-}x_{4,j}^{-}}{1-g^{2}/x_{\bar{6},k}^{+}x_{4,j}^{+}}\, .
\end{eqnarray}
We should however mention that these equations do not 
have any proper solutions for finite values of $M_d$. 
Nevertheless, one may formally proceed and take $M_d$
to infinity. Note that we do {\it not} assume 
any of the original quantum numbers 
$(L;K_1,K_2,K_3,K_4,K_5,K_6,K_7)$
of the asymptotic ansatz to be thermodynamically large.
The remainder of the derivation proceeds again by transforming
\eqref{dressing 2} into Fourier space 
\begin{equation}\label{Fouriertrouble}
0=e^{|t|}\hat{\xi}(t)-\hat{\xi}(t)-\frac{1}{2\pi M_{d}}
\int_{-\infty}^{\infty}\D t' \sum_{j=1}^{K_4}e^{-it'u_{j}} e^{|t|} 
\left(\hat{K}(2gt,2gt')+\hat{H}(2gt,2gt')\right)e^{|t'|/2}
\end{equation}  
where $\hat{K}(t,t')$ and $\hat{H}(t,t')$ are given by \eqref{Khat} and \eqref{Hhat} 
and $\hat{\xi}(t)$ is the density of the dressing roots.
If we do not assume the distribution of roots to be an even function, we end up with
two dressing densities, for $t>0$ and $t<0$ respectively
\begin{equation}\label{xi t>0}
\hat{\xi}(\pm t)=\frac{-2g^{2}t}{M_{d}}\frac{1}{e^{t}-1}
\int_{0}^{\infty}\D t' e^{-t'/2}\Big(\sum_{j=1}^{K_4}e^{\pm it'u_{j}}\hat{K}_{m}(2gt,2gt')
+\sum_{j=1}^{K_4} e^{\mp it'u_{j}}\hat{H}_{m}(2gt,2gt')\Big),
\end{equation}
with $\hat{K}_{m}$ given by (\ref{Km}) and
\begin{equation}\label{Hm}
\hat{H}_{m}(t,t')=\frac{\BesselJ_{0}(t)\BesselJ_{1}(t')+\BesselJ_{0}(t')\BesselJ_{1}(t)}{t+t'}\,.
\end{equation}
Due to symmetry the second set of 
dressing roots can be treated in the same way. 
After elimination of the auxiliary densities in \eqref{dressing main} one 
finds 
\begin{equation}\label{2itheta}
2i\theta_k(x_{1},\ldots,x_{M})=
\arg \sigma^2_k(x_{1},\ldots,x_{M})=\nonumber
\end{equation}
\begin{equation}
2g^{2}\int_{-\infty}^{\infty}\D t~e^{-\frac{|t|}{2}}e^{itu_{k}}\int_{-\infty}^{\infty}\D t'~e^{-\frac{|t'|}{2}}\sum_{j=1}^{M}e^{it'u_{j}}\left(\hat{K}_{d}(2gt',2gt)-\hat{K}_{d}(2gt,2gt') \right)
\end{equation}
with the dressing kernel given by \eqref{Kd}. Comparing the last equation
with \eqref{dressingfactor} we notice that the dressing phase 
factorizes as promised in \eqref{factor}.

The above procedure is suggestive, but currently plagued
by the following problems: 

\begin{itemize}

\item{ There is no solution to the
equations \eqref{FULL SET1},\eqref{FULL SET4}
relating to the dressing roots when their number 
$M_d$ is finite. Therefore expressions such as 
\eqref{Fouriertrouble}, where we employ $M_d$ as an infrared regulator, are somewhat ill-defined. This is in contradistinction 
to our treatment of $\fldF^L$ where the equations certainly make
sense at finite $L$. }

\item{ The scattering pattern of 
\eqref{MAIN} - \eqref{FULL SET4} corresponds to
a very curious ``Dynkin diagram'' whose Lie-algebraic
origin is very questionable, see Figure 2. Furthermore,
the occupation (filling) numbers appear to be inconsistent 
with a standard nested Bethe ansatz. It would have to
be shown how such filling numbers can be obtained from
the non-compact nature of the underlying system.}

\item{The ansatz \eqref{MAIN} - \eqref{FULL SET4} 
leads to Bethe equations which are at $M_d=\infty$ strictly identical 
to the original ones with the dressing factor. So the problem of the
asymptotic character of the equations of \cite{LongRange} 
is not improved, and finite size effects are still not  properly
implemented. }

\item{ Finally, it is completely unclear which
``hidden" excitations are supposed to be scattering according to 
\eqref{MAIN} - \eqref{FULL SET4}. It is certainly not clear what,
if anything, is being diagonalized by these equations.
The equations might however turn out to be useful for proving
the crossing invariance of the dressing factor at finite
values of the coupling constant.}

\end{itemize}

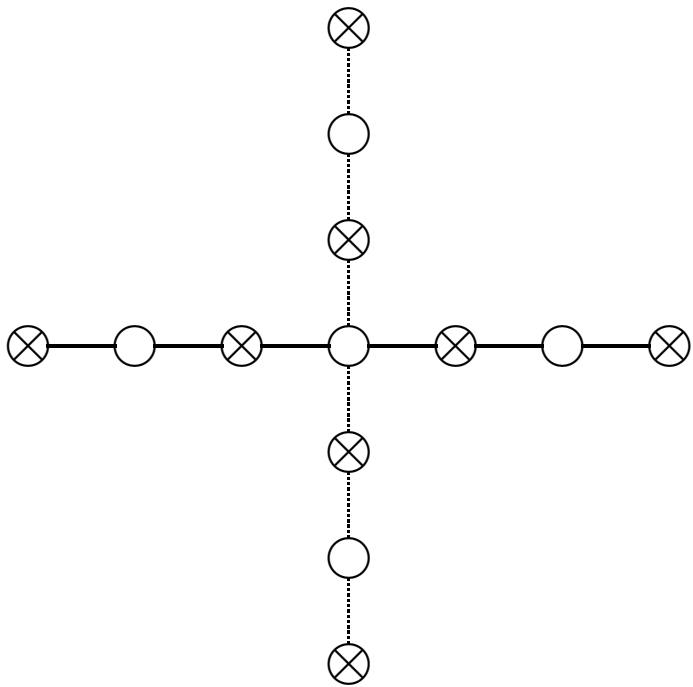
\begin{figure}[t]\label{fig:Cross}
\begin{minipage}{260pt}
\begin{picture}(260,250)(-110,-130)
\setlength{\unitlength}{1pt}
\small\thicklines
\multiput(120, 40)(00, 40){3}{\circle{15}}
\multiput(115, 45)(00, 80){2}{\line(1,-1){10}}
\multiput(115, 35)(00, 80){2}{\line(1 ,1){10}}
\dottedline{2}(120,08)(120,32)
\dottedline{2}(120,48)(120,72)
\dottedline{2}(120,88)(120,112)
%\multiput(120, 07)(00, 40){3}{\line(0, 1){26}}
%
\multiput(  0, 00)(40, 00){7}{\circle{15}}
\multiput( -5, 05)(80, 00){4}{\line(1,-1){10}}
\multiput( -5, -5)(80, 00){4}{\line(1,1){10}}
\multiput(  7, 00)(40, 00){6}{\line(1,0){26}}
\multiput(120,-40)(00,-40){3}{\circle{15}}
\multiput(115,-35)(00,-80){2}{\line(1,-1){10}}
\multiput(115,-45)(00,-80){2}{\line(1 ,1){10}}
\dottedline{2}(120,-08)(120,-32)
\dottedline{2}(120,-48)(120,-72)
\dottedline{2}(120,-88)(120,-112)
%\multiput(120,-07)(00,-40){3}{\line(0,-1){26}}
%
\end{picture}
\end{minipage}
\caption{Schematic scattering notation for the dressed 
asymptotic Bethe ansatz equations \eqref{MAIN} - \eqref{FULL SET4}.
The horizontal branch represents the Dynkin diagram of $\mathfrak{psu}(2,2|4)$
and the scattering of elementary magnons. The vertical branch 
(dotted lines)
indicates the emulation of the dressing phase by the scattering of the dressing roots among
themself and the momentum carrying main node. Note that the auxiliary ``dressing'' roots do {\it{not}} scatter
with any of the auxiliary ``nesting'' roots.}
\end{figure}

%
%%%%%%%%%%%%%%%%%%%%%%%%%%%%%%%%%%%%%%%%%%%%%%%%%%%%%%%%%%%%%%%%%%

%%%%%%%%%%%%%%%%%%%%%%%%%%%%%%%%%%%%%%%%%%%%%%%%%%%%%%%%%%%%
\end{document}